\documentclass{article}
\usepackage{amssymb,amsmath,amsfonts,graphicx,hyperref}

\input{tcilatex}
\begin{document}

\title{\bf Turbulence and Shock-Waves in Crowd Dynamics}
\author{Vladimir G. Ivancevic and Darryn J. Reid \\
Land Operations Division\\
Defence Science \& Technology Organisation}\date{}
\maketitle

\begin{abstract}
In this paper we analyze crowd turbulence from both classical and quantum
perspective. We analyze various crowd waves and collisions using crowd
macroscopic wave function. In particular, we will show that nonlinear Schr%
\"{o}dinger (NLS) equation is fundamental for quantum turbulence, while its
closed-form solutions include shock-waves, solitons and rogue waves, as well
as planar de Broglie's waves. We start by modeling various crowd flows using
classical fluid dynamics, based on Navier--Stokes equations. Then, we model
turbulent crowd flows using quantum turbulence in Bose-Einstein
condensation, based on modified NLS equation.\\

\noindent \textbf{Keywords:} Crowd behavior dynamics, classical and quantum
turbulence, shock waves, solitons and rogue waves
\end{abstract}

\tableofcontents

\section{Introduction}

Massive crowd movements can be today precisely observed from satellites. All
that one can see is physical movement of the crowd. Therefore, all involved
psychology of individual crowd agents (and their groups within the crowd):
cognitive, motivational and emotional, as well as its global sociology, is
only a non-transparent input (a hidden initial switch) to the fully
observable crowd physics \cite{NlDyn1,NlDyn2,TopDual,IvResonan}. About a
decade ago, D. Helbing discovered a phenomenon called \emph{crowd turbulence}
(see \cite{HelbingPRE1,HelbingNature,HelbingPRL,HelbingPRE,HelbingACS}),
depicting crowd disasters caused by the panic stampede that can occur at
high pedestrian densities and which is a serious concern during various
disasters (bushfires, tornados, earthquakes), as well as mass events like
soccer championship games or annual pilgrimage in Mecca.

The adaptive, wave-form, nonlinear and stochastic crowd dynamics has been
modeled using (an adaptive form of) \emph{nonlinear Schr\"{o}dinger equation}%
\footnote{%
The most important case of nonlinear Schr\"{o}dinger equation is the \textit{%
cubic NLS}
\begin{equation*}
\mathrm{i}\partial _{t}\psi =-\frac{1}{2}\Delta \psi \pm |\psi |^{2}\psi ,
\end{equation*}%
with the cubic nonlinearity $|\psi |^{2}\psi $. The sign $+$ in the cubic
NLS represents \textit{defocusing NLS}, while the $-$ sign represents
\textit{focusing NLS}. This extraordinarily rich nonlinear PDE represents a
fully integrable Hamiltonian system which is traditionally studied on
Euclidean domains $\mathbb{R}^{n}$ (but other domains, like circle, torus or
hypersphere, are also studied) -- and allows for a `zoo' of various
wave-like solutions (to be analyzed later). Terms \textit{sub--critical},
\textit{critical}, and \textit{super--critical} are frequently used to
denote a significant transition in the behavior of a particular equation
with respect to a specified regularity class (or conserved quantity).
Typically, sub--critical equations behave in an approximately linear manner,
supercritical equation behave in a highly nonlinear manner, and critical
equations are very finely balanced between the two. Occasionally one also
discusses the sub--criticality, criticality, or super--criticality of
regularities with respect to other symmetries than scaling, such as Galilean
invariance or Lorentz invariance. For survey of recent advances in nonlinear
wave equations based on their criticality, see \cite{tao:cbms}.} (NLS), also
called \emph{Gross--Pitaevskii equation}\footnote{%
NLS or GP is also similar in form to the Ginzburg--Landau equation, a
mathematical theory used to model superconductivity.}\emph{\ }(GP), defining
the time-dependent complex-valued \emph{macroscopic wave function} $\psi
=\psi (x,t)$, whose absolute square $|\psi (x,t)|^{2}$ represents the
\textit{crowd density function}. In natural quantum units ($\hbar=1,~m=1$),
our NLS equation reads:
\begin{equation}
\mathrm{i}\partial _{t}\psi =-\frac{1}{2}\partial _{xx}\psi +V|\psi
|^{2}\psi ,\qquad (\mathrm{i}=\sqrt{-1};\quad \text{with \ }\partial
_{z}\psi =\frac{\partial \psi }{\partial z}),\qquad  \label{nlsGen}
\end{equation}%
where $V=V(w,x)$ denotes the adaptive heat potential (trained by either
Hebbian or Levenberg--Marquardt learning). Physically, the NLS equation (\ref%
{nlsGen}) describes a nonlinear wave in a quantum matter (such as
Bose-Einstein condensates).
\begin{figure}[h]
\centerline{\includegraphics[width=14cm]{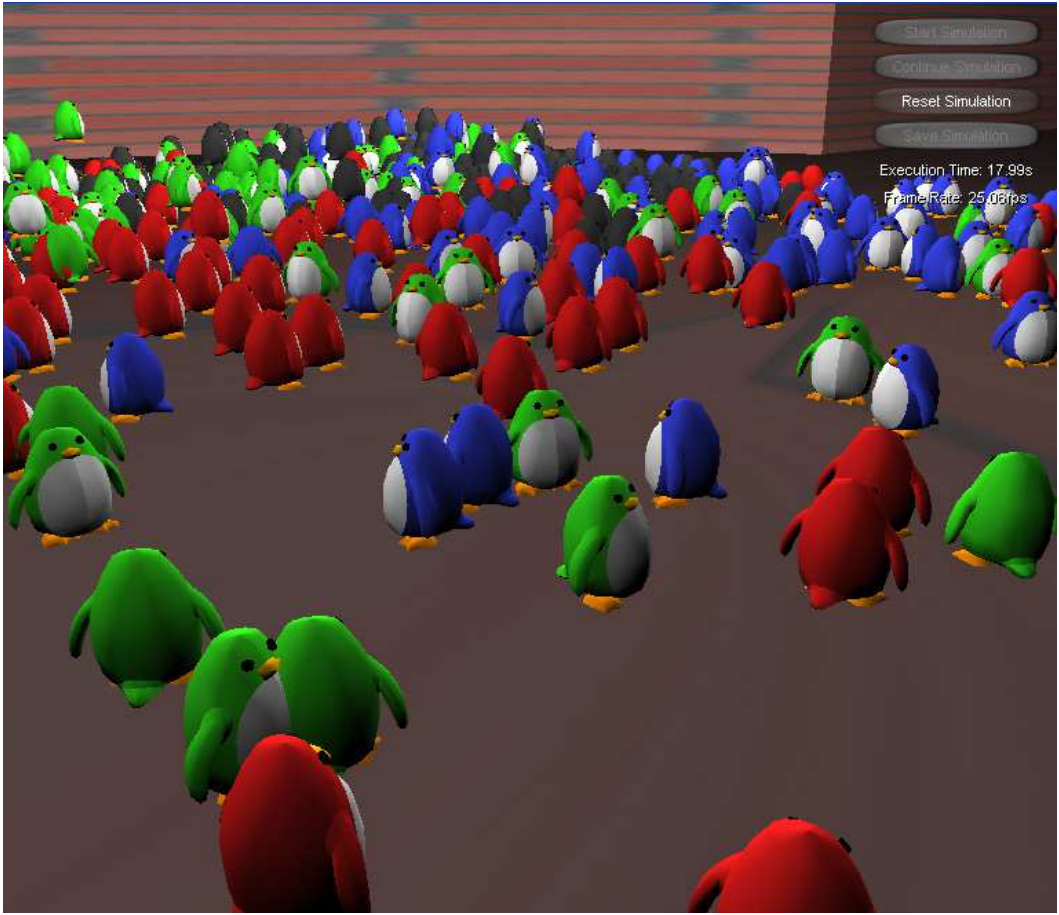}}
\caption{Simulating random-mixing behavior of four crowds in a confined environment.}
\label{DaVinci}
\end{figure}

For general crowd simulation, we recently proposed two NLS-based approaches
in \cite{IvResonan}, each representing a \emph{quantum neural network} \cite%
{QnnBk}:

\begin{description}
\item[Weak coupling approach:] A looser (and more abstract) but higher-dimensional
approach consisting of an $n-$dimensional set of NLS equations:
\begin{equation}
\mathrm{i}\partial _{t}\psi _{i}=-\frac{1}{2}\partial _{xx}\psi _{i}+V|\psi
_{i}|^{2}\psi _{i},\qquad (i=1,...,n),  \label{weakNLS}
\end{equation}%
which self-organize in a common adaptive potential $V=V(w,x)$. Here, the
squared amplitude $|\psi |^{2}$ is the \textit{condensate density. }The
potential $V(w,x)$ includes synaptic weights $w_{k}$, which iteratively
update according to the standard Hebbian rule:
\begin{equation*}
\dot{w}_{i}=-w_{i}+c|\psi |g_{i}|\psi |,\qquad
V(w,x)=\sum_{i=1}^{n}w_{i}g_{i},
\end{equation*}%
where $c$ is the learning rate parameter and $g_{i}$ are Gaussian kernel
functions with means $x_{i}$ and standard deviations normalized to unity.
The system (\ref{weakNLS}) was numerically solved using the Method of Lines
(combined with the fast adaptive Runge-Kutta-Fehlberg integrator with
Cash-Karp accelerator). Its solution\footnote{%
Crowd simulations were based on the following data:
\par
\begin{enumerate}
\item Target function used for this case is: $f=2\sin (20\pi t)$. An
infinitesimal tracking function used is: $h=-(PDF\cdot dx/x)$.
\par
\item Gaussian kernel functions are defined as: $g_{i}=\exp
[-(v-m_{i})^{2}], $ ~where $v=f-h$, while $m_{i}$ are random Gaussian means.
\par
\item Potential field update is given by the scalar product: $%
V_{i+1}=V_{i}+w_{i}\cdot g_{i}$.
\par
\item The complex plane $\mathbb{C}$ was embedded in the 3D graphics
environment (both urban and bush) with 3D collision dynamics.
\end{enumerate}
} represents time evolution in the complex plane $\mathbb{C}$ of $n$
cooperative groups, each consisting of $m$ agents of $SE(2)$-kinematic type.%
\footnote{%
The Euclidean motion group $SE(2)\equiv SO(2)\times \mathbb{R}$ is a set of
all $3\times 3-$ matrices of the form:
\begin{equation*}
\left[
\begin{array}{ccc}
\cos \theta & \sin \theta & x \\
-\sin \theta & \cos \theta & y \\
0 & 0 & 1%
\end{array}%
\right] ,
\end{equation*}%
including both rigid translations (i.e., Cartesian $x,y-$coordinates) and
rotation matrix $\left[
\begin{array}{cc}
\cos \theta & \sin \theta \\
-\sin \theta & \cos \theta%
\end{array}%
\right] $ in Euclidean plane $\mathbb{R}^{2}\approx \mathbb{C}$ (see \cite%
{GaneshSprBig,GaneshADG}).} In this approach, each individual line (or
kinematic trajectory), defines a velocity controller for a single agent. The
total number of agents, as well as number of groups, is limited only by the
available computation power.

\item[Strong coupling approach:] A pair of strongly-coupled NLS equations
with Hebbian learning:
\begin{eqnarray}
{\text{BLUE}} &:&\quad \mathrm{i}\partial _{t}\psi _{\mathrm{B}}~=~-\frac{a_{%
\mathrm{B}}}{2}|\phi _{\mathrm{R}}|^{2}\partial _{xx}\psi _{\mathrm{B}%
}+V|\psi _{\mathrm{B}}|^{2}\psi _{\mathrm{B}},  \notag \\
{\text{RED}} &:&\quad \mathrm{i}\partial _{t}\phi _{\mathrm{R}}~=~-\frac{b_{%
\mathrm{R}}}{2}|\psi _{\mathrm{B}}|^{2}\,\partial _{xx}\phi _{\mathrm{R}%
}+V|\phi _{\mathrm{R}}|^{2}\phi _{\mathrm{R}},  \label{strongNLS} \\
{\text{HEBB}} &:&\quad \dot{w}_{i}~=~-w_{i}+c_{\mathrm{H}}|\psi _{\mathrm{B}%
}|g_{i}|\phi _{\mathrm{R}}|,\qquad V~=~\sum_{i=1}^{n}w_{i}g_{i}.  \notag
\end{eqnarray}%
Here, $a_{\mathrm{B}},b_{\mathrm{R}},c_{\mathrm{H}}$ are parameters related
to Red, Blue and Hebb, equations respectively. This is a bidirectional,
spatiotemporal, complex-valued associative memory machine, generalizing
Lanchester and Lotka-Volterra predator-prey dynamical systems, as well as
Hopfield, Kosko and Grossberg models of neural networks (see, e.g. \cite%
{NeuFuz}).
\end{description}

In this paper we will analyze crowd turbulence (from both classical and
quantum perspective) as well as various crowd waves and collisions. In
particular, we will show that NLS equation (\ref{nlsGen}) is fundamental for
quantum turbulence, while its closed-form solutions include shock-waves,
solitons and rogue waves. Firstly, we will model various crowd flows using
classical fluid dynamics, based on Navier--Stokes equations. Then, we will
model turbulent crowd flows using quantum turbulence in Bose-Einstein
condensation, based on modified NLS equation.

\section{Classical Approach to Crowd Turbulence}

In this section we model various crowd flows using models from classical
fluid dynamics, based on Navier--Stokes partial differential equations
(PDEs).

\subsection{Classical Turbulence and Crowd Flows}

Turbulence has long been one of the great mysteries in nature, with
discussion dating back to the era of Leonardo da Vinci. He observed the
turbulent flow of water and drew pictures showing that turbulence has a
structure comprised of vortices of different sizes (Figure \ref{DaVinci}).
After Leonardo, turbulence has been intensely studied in a number of fields,
but it is still far from completely understood. This is primarily because
turbulence is a strongly nonlinear dynamical phenomenon.
\begin{figure}[h]
\centerline{\includegraphics[width=10cm]{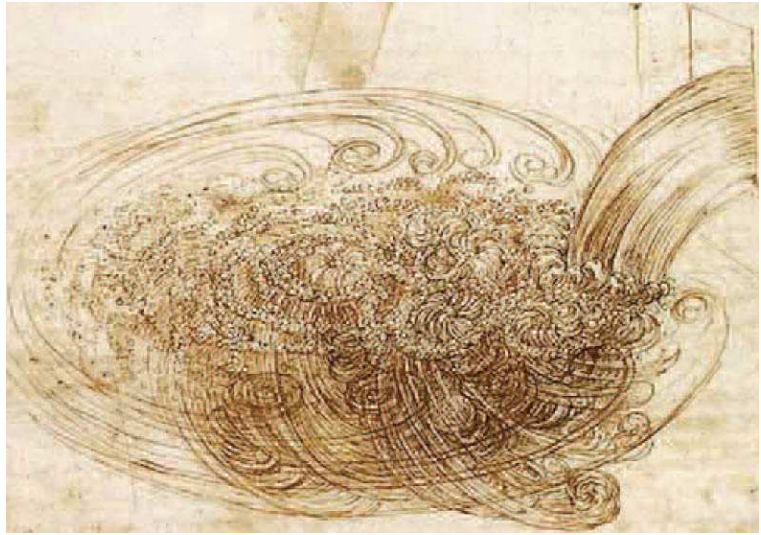}}
\caption{Sketch of turbulence by Leonardo da Vinci.}
\label{DaVinci}
\end{figure}

\emph{Turbulent flow} is a fluid flow regime characterized by low momentum
diffusion, high momentum convection, and rapid variation of pressure and
velocity in space and time. Flow that is not turbulent is called \textit{%
laminar flow}. Also, recall that the \textit{Reynolds number} $Re$
characterizes whether flow conditions lead to laminar or turbulent flow. The
structure of turbulent flow was first described by A. Kolmogorov. Consider
the flow of water over a simple smooth object, such as a sphere. At very low
speeds the flow is laminar, i.e., the flow is locally smooth (though it may
involve vortices on a large scale). As the speed increases, at some point
the transition is made to turbulent (or, chaotic) flow. In turbulent flow,
unsteady \emph{vortices}\footnote{%
Vortex can be any circular or rotary flow that possesses vorticity. Vortex
represents a spiral whirling motion (i.e., a spinning turbulent flow) with
closed streamlines. The shape of media or mass rotating rapidly around a
center forms a vortex. It is a flow involving rotation about an arbitrary
axis and can be described by the vector curl operator.
\par
In the atmospheric sciences, vorticity is a property that characterizes
large--scale rotation of air masses. Since the atmospheric circulation is
nearly horizontal, the 3D vorticity is nearly vertical, and it is common to
use the vertical component as a scalar vorticity.
\par
A vortex can be seen in the spiraling motion of air or liquid around a
center of rotation. Circular current of water of conflicting tides form
vortex shapes. Turbulent flow makes many vortices. A good example of a
vortex is the atmospheric phenomenon of a whirlwind or a \textit{tornado}.
This whirling air mass mostly takes the form of a helix, column, or spiral.
Tornadoes develop from severe thunderstorms, usually spawned from squall
lines and \textit{supercell thunderstorms}, though they sometimes happen as
a result of a \textit{hurricane}. (A hurricane is a much larger, swirling
body of clouds produced by evaporating warm ocean water and influenced by
the Earth's rotation. In particular, polar vortex is a persistent,
large--scale cyclone centered near the Earth's poles, in the middle and
upper troposphere and the stratosphere. Similar, but far greater, vortices
are also seen on other planets, such as the permanent Great Red Spot on
Jupiter and the intermittent Great Dark Spot on Neptune.) Another example is
a meso-vortex on the scale of a few miles (smaller than a hurricane but
larger than a tornado). On a much smaller scale, a vortex is usually formed
as water goes down a drain, as in a sink or a toilet. This occurs in water
as the revolving mass forms a whirlpool. (A whirlpool is a swirling body of
water produced by ocean tides or by a hole underneath the vortex, where
water drains out, as in a bathtub.) This whirlpool is caused by water
flowing out of a small opening in the bottom of a basin or reservoir. This
swirling flow structure within a region of fluid flow opens downward from
the water surface. In the hydrodynamic interpretation of the behavior of
electromagnetic fields, the acceleration of electric fluid in a particular
direction creates a positive vortex of magnetic fluid. This in turn creates
around itself a corresponding negative vortex of electric fluid.} appear on
many scales and interact with each other. Drag due to boundary layer skin
friction increases. The structure and location of boundary layer separation
often changes, sometimes resulting in a reduction of overall drag.

Applied to crowd dynamics, \emph{vorticity} $\mathbf{\omega }=\mathbf{\omega
}(\mathbf{x},t)$ is defined as the circulation per unit area at a point in
the crowd flow field, that is as the curl of the the crowd flow velocity:~$%
\mathbf{\omega }=\nabla \times \mathbf{u}$. It is a vector quantity, whose
direction is approximately along the axis of the swirl. The movement of
a crowd flow can be said to be vortical if the fluid moves around in a
circle, or in a helix, or if it tends to spin around some axis. Such motion
can also be called \emph{solenoidal}.

Because laminar--turbulent transition in crowd dynamics is governed by
Reynolds number, the same transition occurs if the size of the crowd is
gradually increased, or the viscosity of the crowd is decreased, or if the
density of the crowd is increased.

In particular, in a turbulent crowd flow, there is a range of scales of the
crowd flow motions, called \emph{eddies}. A single packet of crowd flow
moving with a bulk velocity is called an `eddy'. The size of the largest
scales (eddies) are set by the overall geometry of the crowd flow.\footnote{%
For comparison, in an industrial smoke-stack, the largest scales of fluid
motion are as big as the diameter of the stack itself. The size of the
smallest scales is set by $Re$. As $Re$ increases, smaller and smaller
scales of the flow are visible. In the smoke-stack, the smoke may appear to
have many very small bumps or eddies, in addition to large bulky eddies. In
this sense, $Re$ is an indicator of the range of scales in the flow. The
higher the Reynolds number, the greater the range of scales.}

Such turbulent crowd flow shows characteristic statistical behavior (compare
with \cite{Kolmogorov41a, Kolmogorov41b}). For simplicity, we will assume a
steady state of fully developed turbulence of an incompressible classical
crowd flow. Energy is injected into the crowd flow at a rate $\varepsilon $
in an energy-containing range. In an inertial range, this crowd energy is
transferred to smaller length scales without dissipation. In this range, the
crowd is locally homogeneous and isotropic, leading to energy spectral
statistics described by the \textit{Kolmogorov law,}\footnote{%
The Kolmogorov spectrum has been confirmed experimentally and numerically in
fluid turbulence at high Reynolds numbers.}
\begin{equation}
E(k)=C\,\varepsilon ^{2/3}\,k^{-5/3}.  \label{eq-Kolmogorov}
\end{equation}%
The crowd energy spectrum $E(k)$ is defined by $E=\int E(k)\,d\mathbf{k}$,
where $E$ is the kinetic energy of the crowd per unit mass and $k$ is the
crowd wave-number from the Fourier transform of the velocity field (compare
with tha last subsection on quantum crowd waves). The spectrum of equation (%
\ref{eq-Kolmogorov}) is derived by assuming that $E(k)$ is locally
determined only by the crowd energy flux $\varepsilon $ and by $k$. The
crowd energy transferred to smaller scales is dissipated at the Kolmogorov
wave-number $k_{K}=(\epsilon /\nu ^{3})^{1/4}$ in an energy-dissipative
range via the viscosity of the crowd flow at a dissipation rate $\varepsilon
$ in equation (\ref{eq-Kolmogorov}), which is equal to the crowd energy flux
$\Phi $ in the inertial range. The Kolmogorov constant $C$ is a
dimensionless parameter of order unity. The inertial range is thought to be
sustained by a self-similar \emph{Richardson cascade} in which large crowd
eddies break up into smaller eddies through crowd vortex reconnections.

In order for two crowd flows to be similar they must have the same geometry
and equal Reynolds numbers. When comparing crowd flow behavior at homologous
points in a crowd model and a full--scale crowd flow, we have ~$Re^{\ast
}=Re $,~ where quantities marked with $^{\ast }$ concern the flow around the
crowd model and the other the real crowd flow.

\subsection{Navier-Stokes Crowd Fluids}

Fluid dynamicists believe that Navier--Stokes PDEs accurately describe
turbulence (see, e.g. \cite{Frisch}). Therefore, we can assume that viscous
crowd flows evolve according to nonlinear Navier--Stokes equations\footnote{%
Navier--Stokes equations, named after C.L. Navier and G.G. Stokes, are a set
of PDEs that describe the motion of liquids and gases, based on the fact
that changes in momentum of the particles of a fluid are the product of
changes in pressure and dissipative viscous forces acting inside the fluid.
These viscous forces originate in molecular interactions and dictate how
viscous a fluid is, so the Navier--Stokes PDEs represent a dynamical
statement of the balance of forces acting at any given region of the fluid.
They describe the physics of a large number of phenomena of academic and
economic interest (they are useful to model weather, ocean currents, water
flow in a pipe, motion of stars inside a galaxy, flow around an airfoil
(wing); they are also used in the design of aircraft and cars, the study of
blood flow, the design of power stations, the analysis of the effects of
pollution, etc).}
\begin{equation}
\mathbf{\dot{u}}+\mathbf{u}\cdot \mathbf{\nabla }\mathbf{u}+{\nabla p}/\rho
=\nu \Delta \mathbf{u}+\mathbf{f},  \label{NavSt}
\end{equation}%
where $\mathbf{u}=\mathbf{u}(\mathbf{x},t)$~ is the 3D velocity of a crowd
flow , $\mathbf{\dot{u}\equiv \partial }_{t}\mathbf{u}$~ is the 3D
acceleration of a crowd flow , $p=p(\mathbf{x},t)$ is the crowd pressure
field, while $\mathbf{f}=\mathbf{f}(\mathbf{x},t)$ is the external nonlinear
energy source to the crowd, while $\rho ,\nu $ are the crowd flow density
and viscosity coefficient, respectively. Such a crowd flow can be
characterized by the ratio of the second term on the left-hand side of
equation (\ref{NavSt}), $\mathbf{u}\cdot \mathbf{\nabla }\mathbf{u}$,
referred to as the crowd inertial term, and the second term on the
right-hand side, $\nu \Delta \mathbf{u}$, that we call the crowd viscous
term. This ratio defines the \textit{Reynolds number}\footnote{%
Reynold's number $\mathit{Re}$ is the most important dimensionless number in
fluid dynamics and provides a criterion for determining \textit{dynamical
similarity}. Where two similar objects in perhaps different fluids with
possibly different flow--rates have similar fluid flow around them, they are
said to be dynamically similar. $Re$ is the ratio of inertial forces to
viscous forces and is used for determining whether a flow will be \emph{%
laminar} or \emph{turbulent}. Laminar flow occurs at low Reynolds numbers,
where viscous forces are dominant, and is characterized by smooth, constant
fluid motion, while turbulent flow, on the other hand, occurs at high $Re$s
and is dominated by inertial forces, producing random eddies, vortices and
other flow fluctuations. The transition between laminar and turbulent flow
is often indicated by a critical Reynold's number ($Re_{crit}$), which
depends on the exact flow configuration and must be determined
experimentally. Within a certain range around this point there is a region
of gradual transition where the flow is neither fully laminar nor fully
turbulent, and predictions of fluid behavior can be difficult.} $Re=\bar{v}%
D/\nu $, where $\bar{v}$ and $D$ are a characteristic velocity and length
scale, respectively. When $\bar{v}$ increases and the Reynolds number $Re$
exceeds a critical value, the crowd changes from a \textit{laminar} to a
\textit{turbulent} state, in which the crowd flow is complicated and
contains eddies.\footnote{%
The first demonstration of the existence of an unstable recurrent pattern in
a turbulent hydrodynamic flow was performed in \cite{Kawahara}, using the
full numerical simulation, a 15,422--dimensional discretization of the 3D
Plane Couette turbulence at the Reynold's number $Re=400$. The authors found
an important unstable spatiotemporally--periodic solution, a single unstable
recurrent pattern.} To simplify the problem, we can impose to $\mathbf{f}$
the so--called \emph{Reynolds condition}, $\langle \mathbf{f}\cdot \mathbf{u}%
\rangle =\varepsilon $, where $\varepsilon $ is the average rate of energy
injection.

In mechanical Lie algebra terms (see Appendix), the Navier--Stokes PDE (\ref%
{NavSt}) can be written:%
\begin{equation*}
\mathbf{\dot{\omega}}=-[\mathbf{u},\mathbf{\omega }]+\text{curl }\mathbf{f}%
+\nu \Delta \mathbf{\omega }.
\end{equation*}

On the other hand, the \textit{Euler equation} for ideal crowd flows,

\begin{equation}
\mathbf{\dot{u}}+\mathbf{u}\cdot \mathbf{\nabla }\mathbf{u}+{\nabla p}/\rho
=0,  \label{EulerFluid}
\end{equation}%
reads in Lie algebra terms,
\begin{equation*}
\mathbf{\dot{\omega}}=-[\mathbf{u},\mathbf{\omega }],\qquad \mathbf{\omega }=%
\text{curl }\mathbf{u}.
\end{equation*}%
Equation (\ref{EulerFluid})\ is related to the Navier--Stokes PDE (\ref%
{NavSt}) in the same way as the classical Euler equation of a rigid body
(with a fixed point, see Appendix),
\begin{equation*}
\mathbf{\dot{\pi}}=\mathbf{\pi }\times \mathbf{\omega },
\end{equation*}%
is associated to a more general equation, involving friction and external
angular momentum \cite{ArnoldMech,ArnoldHidr}%
\begin{equation}
\mathbf{\dot{\pi}}=\mathbf{\pi }\times \mathbf{\omega }+\mathbf{F}-\nu
\mathbf{\pi },  \label{EulerBody}
\end{equation}%
the `friction operator' $\nu $ is symmetric and positive definite. The
distributed mass force $\mathbf{f}$, which appeared in the Navier-Stokes
equation (\ref{NavSt}), is similar to the external angular momentum $\mathbf{%
F}$, and it is the origin of the crowd flow motion. The viscous friction $%
\nu \Delta \mathbf{u}$ is analogous to the term $-\nu \mathbf{\pi }$ in (\ref%
{EulerBody}) slowing the rigid body motion. The similarity becomes
especially noticeable if one rewrites the equations in components in the
eigenbasis of the friction operator.

For example, for the Navier-Stokes equation with periodic boundary
conditions one can expand the crowd vorticity field and the force $\mathbf{f}
$ into the ordinary Fourier series. The equations in both of the cases have
the form of \textit{Galerkin approximation}:%
\begin{equation}
\dot{x}_{i}=\sum_{j,k}a_{ijk}x_{j}x_{k}+\sum_{i}f_{i}-\nu _{i}x_{i}.
\label{Galerkin}
\end{equation}%
The first term corresponds to the Euler equation (\ref{EulerFluid})\ and
describes the inertial crowd motion. It follows from the properties of (\ref%
{EulerFluid}) that the divergence of this term is equal to zero.
Furthermore, the Euler equation of an ideal crowd flow in any dimension, as
well as that of a rigid body, has a quadratic positive definite first
integral, the kinetic energy. Therefore, for $\mathbf{f}=\mathbf{u}=0,$ the
vector field on the right-hand side of (\ref{Galerkin}) is tangent to
certain ellipsoids centered at the crowd origin. This implies that during
the crowd evolution defined by this equation there is neither growth nor
decay of solutions.

The term $-\nu _{i}x_{i}$ in (\ref{Galerkin}), corresponding to the crowd
friction, dominates over the constant `crowd pumping' $\mathbf{f}$ when
considered sufficiently far away from the crowd origin. Hence, in that
remote crowd region, the crowd motion is directed towards the origin, and an
infinite growth of solutions is impossible. Also, since the `crowd pumping' $%
\mathbf{f}$ pushes a phase point out of any neighborhood of the origin,
while the friction returns it from a distance, crowd motion in the system of
a rigid body (\ref{EulerBody}) approaches an intermediate regime-attractor.
For instance, this crowd attractor can be a stable stagnation point or a
periodic crowd motion, while for sufficiently high dimension of the phase
space it can appear to be a \textit{chaotic motion} sensitive to the initial
conditions.

Here we recall that \textit{chaos theory,} of which turbulence is the most
extreme form, started in 1963, when E. Lorenz from MIT took the
Navier--Stokes PDEs (\ref{NavSt}) and reduced them into three first--order
coupled nonlinear ODEs, to demonstrate the idea of sensitive dependence upon
initial conditions and associated \textit{chaotic behavior}. The 3D \emph{%
phase--portrait} of the Lorenz system (\ref{LorenzSys}) shows the celebrated
`\textit{Lorenz mask}', a special type of \textit{fractal attractor}.%
\footnote{%
The Lorenz reduced system of nonlinear ODEs
\begin{equation}
\dot{x}=a(y-x),\qquad \dot{y}=bx-y-xz,\qquad \dot{z}=xy-cz,
\label{LorenzSys}
\end{equation}%
where $x$, $y$ and $z$ are dynamical variables, constituting the 3D \emph{%
phase--space} of the \textit{Lorenz flow}; and $a$, $b$ and $c$ are the
parameters of the system. Originally, Lorenz used this model to describe the
unpredictable behavior of the weather, where $x$ is the rate of convective
overturning (convection is the process by which heat is transferred by a
moving fluid), $y$ is the horizontal temperature overturning, and $z$ is the
vertical temperature overturning; the parameters are: $a\equiv P-$%
proportional to the Prandtl number (ratio of the fluid viscosity of a
substance to its thermal conductivity, usually set at $10$), $b\equiv R-$%
proportional to the Rayleigh number (difference in temperature between the
top and bottom of the system, usually set at $28$), and $c\equiv K-$a number
proportional to the physical proportions of the region under consideration
(width to height ratio of the box which holds the system, usually set at $%
8/3 $). The Lorenz system (\ref{LorenzSys}) has the properties: (i) \emph{%
symmetry}: $(x,y,z)\rightarrow (-x,-y,z)$ for all values of the parameters,
and (ii) the $z-$axis $(x=y=0)$ is \emph{invariant} (i.e., all trajectories
that start on it also end on it).
\par
Today, it is well--known that the Lorenz model is a paradigm for
low--dimensional chaos in dynamical systems and this model or its
modifications are widely investigated in connection with modelling purposes
in meteorology, hydrodynamics, laser physics, superconductivity,
electronics, oil industry, chemical and biological kinetics, etc.
\par
The \textit{Lorenz mask} (3D chaotic attractor) has the following
characteristics: (i) trajectory does not intersect itself in three
dimensions; (ii) trajectory is not periodic or transient; (iii) general form
of the shape does not depend on initial conditions; and (iv) exact sequence
of loops is very sensitive to the initial conditions.}

If the crowd friction (or viscosity) coefficient $\nu $ is high enough, then
the crowd attractor will necessarily be a stable equilibrium position. While
the parameter $\nu $ decreases (i.e., the reciprocal parameter, the Reynolds
number $Re=1/\nu $, increases), bifurcations of the crowd equilibrium are
possible, and the crowd attractor can become a periodic motion and later a
totally `stochastic' one.\footnote{%
The hypothesis that this mechanism is responsible for the phenomenon of
turbulenization of a fluid motion for large Reynolds numbers has been
suggested by many authors. In particular, to normalize the attractor, A.N.
Kolmogorov suggested in 1965 considering the `pumping' proportional to the
same small parameter $\nu $ as viscosity, and he formulated the following
two conjectures \cite{ArnoldHidr}:
\par
\begin{enumerate}
\item The weak conjecture: The maximum of the dimensions of minimal
attractors (attractor is called a minimal attractor if it does not contain
smaller attractors) in the phase space of the Navier-Stokes PDEs (\ref{NavSt}%
) (as well as of their Galerkin approximations (\ref{Galerkin})) grows along
with the Reynolds number $Re=1/\nu $.
\par
\item The strong conjecture: Not only maximum, but also the minimum of the
dimensions of the minimal attractors mentioned above increases with the
Reynolds number $Re.$As far as we know, both of these hypotheses still
remain open.
\end{enumerate}
\par
For the Lorenz system, the role of energy is played by a nonhomogeneous
quadratic function. The instability in the Lorenz model is apparently
stronger than in the Kolmogorov one. One can check how the motion along the
Lorenz strange attractor sensitively depends on the initial conditions,
while for the Kolmogorov model it remains a conjecture. It is proven only
that a stationary flow indeed loses stability as the Reynolds number $Re$
increases.
\par
In 1970 Ruelle and Takens formulated the conjecture that turbulence is the
appearance of attractors with sensitive dependence of motion on the initial
conditions along them in the phase space of the Navier-Stokes equation \cite%
{Ruelle}.}

\subsection{Isovorticial 2D Crowd Flows}

Crowd 2D flow differs sharply from crowd 3D flow. According to V. Arnold, in
the realm of fluid dynamics, the essence of this difference is contained in
the difference in the geometries of the orbits of the co-adjoint
representation (see Appendix) in the two and 3D cases \cite%
{ArnoldMech,ArnoldHidr}. The character of the first, inertia, term in the
Galerkin approximation (\ref{Galerkin}) (of the Navier-Stokes PDEs (\ref%
{NavSt}))\ changes drastically in the passage from 2D crowd flow flows to
three- (or higher-) dimensional ones. The reason lies in the distinctions
among the geometries of the coadjoint orbits of the corresponding
diffeomorphism groups. In other words, in the 2D case the orbits are in some
sense closed and behave like a family of level sets of a function. In the 3D
case the orbits are more complicated; in particular, they are unbounded (and
perhaps dense). The orbits of the coadjoint representation of the group of
diffeomorphisms of a 3D Riemannian manifold can be described in the
following way. Let $v_{1}$ and $v_{2}$ be two vector fields of velocities of
a non-compressible crowd flow in the region $D$. We say that the fields $%
v_{1}$ and $v_{2}$ are isovorticial if there is volume-preserving
diffeomorphism $g:D\rightarrow D$ which carries every closed contour $\gamma
$ in $D$ to a new contour such that the circulation of the first field along
the original contour is equal to the circulation of the second field along
the new contour:
\begin{equation*}
\oint_{\gamma }v_{1}=\oint_{g\gamma }v_{2}
\end{equation*}

The image of an orbit of the co-adjoint representation in a Lie algebra is
the set of vector fields isovorticial to the given vector field. We have the
following law of conservation of circulation \cite{ArnoldMech,ArnoldHidr}:
The circulation of a field of velocities of an ideal crowd flow over a
closed flow contour does not change when the contour is carried by the flow
to a new position.

Now, for simplicity, we will assume that the region $D$ of the crowd flow is
2D and oriented. The metric and orientation give a symplectic structure on $%
D $; the field of crowd velocities has divergence zero and is therefore
Hamiltonian. Therefore, this vector field is given by a Hamiltonian function
(many-valued, in general, if the region $D$ is not simply-connected). The
Hamiltonian function of a field of crowd velocities is called the \textit{%
stream-function}, and is denoted by $\psi $. Thus we have:
\begin{equation*}
v=I\,\text{grad }\psi ,
\end{equation*}%
where $I$ is the operator of clockwise rotation by $90^{\circ }$.

The stream function of the commutator of two crowd vector fields turns out
to be the Jacobian (or the Poisson bracket of Hamiltonian formalism) of the
crowd stream functions of the original vector fields:
\begin{equation*}
\psi _{\lbrack v_{1},v_{2}]}=J(\psi _{1},\psi _{2}).
\end{equation*}

The vorticity (or curl) of a 2D crowd field of velocities is the scalar
function $r$ such that the integral around any oriented crowd region $\sigma
$ in $D$ of the product of $r$ with the oriented area element is equal to
the circulation of the field of crowd velocities around the boundary of $%
\sigma $:
\begin{equation*}
\int_{\sigma }r\,dS=\oint_{\partial \sigma }v.
\end{equation*}%
The crowd vorticity can be now computed in terms of the crowd stream
function as:
\begin{equation*}
r=-\Delta \psi .
\end{equation*}%
In the simply-connected 2D case, isovorticity of crowd vector fields $v_{1}$
and $v_{2}$ means that the functions $r_{1}$ and $r_{2}$ (the vorticities of
these fields) are carried to one another under a suitable volume-preserving
diffeomorphism. Therefore, if two vector fields are in the image of the same
orbit of the co-adjoint representation, then a whole series of functionals
are equal. For example, the integrals of all powers $k$ of the crowd
vorticity are:
\begin{equation*}
\int_{D}r_{1}^{k}\,dS=\int_{D}r_{2}^{k}\,dS.
\end{equation*}%
In particular, Euler's equations of motion of a 2D ideal crowd flow:
\begin{equation*}
\partial _{t}v+v\nabla v=-\,\nabla p,\qquad \text{div }v=0,
\end{equation*}%
have an infinite collection of first integrals. For example, such a first
integral is the integral of any power $k$ of the crowd vorticity of the
field of crowd velocities:
\begin{equation*}
I_{k}=\iint_{D}\left( \partial _{x}v_{2}-\partial _{x}v_{1}\right)
^{k}dx\wedge dy,
\end{equation*}%
where $\wedge $ denotes the antisymmetric wedge product.

Following V. Arnold \cite{ArnoldMech,ArnoldHidr}, we obtain in this way the
following assertions regarding stability of planar stationary crowd flows:

\begin{enumerate}
\item A stationary flow of an ideal crowd flow is distinguished from all
crowd flows isovorticial to it by the fact that it is a conditional extremum
(or critical point) of the crowd kinetic energy.

\item If (i) the indicated critical point is actually an extremum, i.e., a
local conditional maximum or minimum, (ii) it satisfies certain (generally
satisfied) regularity conditions, and (iii) the extremum is non-degenerate
(the second differential is positive- or negative-definite), then the
stationary crowd flow is stable (i.e., is a Lyapunov stable equilibrium
position of Euler's equation).

\item The formula for the second differential of the crowd kinetic energy,
on the tangent space to the manifold of crowd fields which are isovorticial
to a given one, has the following form in the 2D case. Let $D$ be a region
in the Euclidean plane with cartesian coordinates $x$ and $y$. Consider a
stationary crowd flow with stream function $\psi =\psi (x,y).$ Then we have
the quadratic \emph{crowd energy form} $d^{2}H,$ given by
\begin{equation*}
d^{2}H=\frac{1}{2}\iint_{D}(\delta v)^{2}+(\Delta \psi /\nabla \Delta \psi
)(\delta r)^{2}dxdy,
\end{equation*}%
where $\delta v$ is the variation of the field of crowd velocities (i.e., a
vector of the tangent space indicated above), and $\delta r=$ curl $\delta v$%
. We remark here that for a crowd stationary flow, the gradient vectors of
the crowd stream function and its Laplacian are collinear. Therefore the
ratio $\Delta \psi /\nabla \Delta \psi $ makes sense. Furthermore, in a
neighborhood of every point where the gradient of the crowd vorticity is not
zero, the crowd stream function is a function of the vorticity function.
\end{enumerate}

The above assertions lead to the conclusion that the positive or negative
definiteness\footnote{A definite bilinear form is a bilinear form $B(x,x)$ over some vector space $V$ such that the associated quadratic form $Q(x)=B(x,x)$ is definite, that is, has a real value with the same sign (positive or negative) for all non-zero $x$. According to that sign, $B$ is called positive definite or negative definite. If $Q(x)$ takes both positive and negative values, the bilinear form $B(x,x)$ is called indefinite. If $B(x, x) \geq 0$ for all $x$, then $B$ is said to be positive semidefinite. Negative semidefinite bilinear forms are defined similarly.} of the quadratic crowd energy form $d^{2}H$ is a sufficient
condition for stability of the stationary crowd flow under consideration.
The analogous proposition in the linearized fluid dynamics problem is called
Rayleigh's theorem.

Now, consider $N$ crowd vortices with velocity circulations $k_{i},$ $%
(i=1,...,N)$ around them in the Euclidean plane $\mathbb{R}^{2}$. Then the
crowd vorticity at any moment will be concentrated at $N$ points, and the
crowd circulations at each of them will remain constant forever. lt is
convenient to write the evolution of crowd vortices as a dynamical system in
the crowd configuration space for the $N$-vortex system, the space $\mathbb{R%
}^{2N}$ with coordinates $z_{i}=(x_{i},y_{i})$ and symplectic structure $%
\sum_{i}k_{i}dy_{i}\wedge dx_{i}.$ Then the crowd vortex evolution in $%
\mathbb{R}^{2N}$ will be given by the following Kirchhoff--Hamiltonian
system \cite{Kir}:%
\begin{equation*}
k_{i}\dot{x}_{i}=\partial _{y_{i}}H,\qquad k_{i}\dot{y}_{i}=-\partial
_{x_{i}}H.
\end{equation*}

\section{Quantum Approach to Crowd Turbulence}

In this section we model turbulent crowd flows using models of quantum
turbulence in Bose-Einstein condensation, based on modified nonlinear 
Schr\"{o}dinger equation. Here, we want to go beyond classical turbulence as described in the previous section using Navier-Stokes equations. Essentially, we want to achieve two things here: (i) to provide a cleaner (more controllable and repetitive) simulation environment for crowd turbulence; and (ii) the ability to include into this environment a variety of nonlinear waves (e.g., shock-waves, solitons, breathers, rogue waves). From the physical perspective, the so-called `Bose-Einstein condensate' will be our macroscopic quantum crowd superfluid model.

\subsection{Quantum Turbulence}

\begin{figure}[h]
\centerline{\includegraphics[width=9cm]{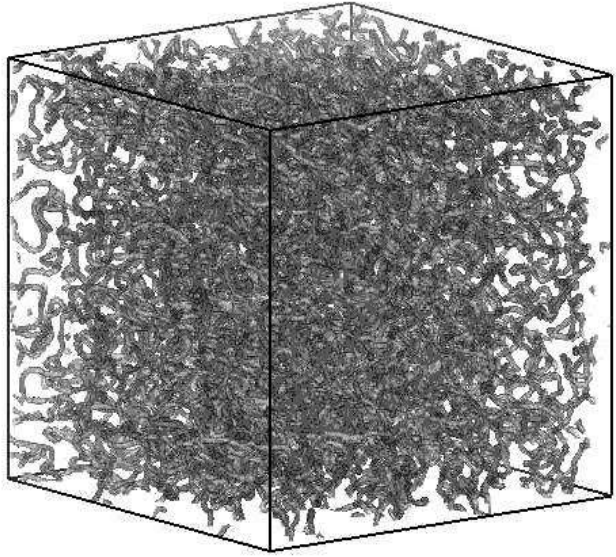}}
\caption{Quantum turbulence (QT) in Bose-Einstein condensation (BEC).}
\label{QT1}
\end{figure}

Quantum turbulence was discovered in superfluid helium ($^{4}$He) in the
1950s, but the field moved in a new direction starting around the mid 1990s
(see \cite{TsVor1,TsVor2}). Briefly, quantum turbulence (see Figure \ref{QT1}%
) is comprised of quantized vortices that are definite topological defects
arising from the order parameter appearing in \emph{Bose-Einstein
condensation} (BEC).\footnote{%
A Bose--Einstein condensate (BEC) is a state of matter of a dilute gas of
weakly interacting bosons confined in an external potential and cooled to
temperatures very near absolute zero. It is the most common example of
\textit{quantum media}.} Hence quantum turbulence is expected to yield a
simpler model of turbulence than does classical turbulence based on the
Navier--Stokes PDE (\ref{NavSt}).

Bose-Einstein condensation is often considered to be a \textit{macroscopic
quantum phenomenon} (see, e.g. \cite{Pitaevskii}). This is because bosons
occupy the same single-particle ground state below a critical temperature
through Bose-Einstein condensation to form a \emph{macroscopic wave function}
(order parameter) extending over the entire system. As a direct result of
the formation of a macroscopic wave function, quantized vortices appear in
the Bose-condensed system. A \emph{quantized vortex}\footnote{%
The studies of quantized vortices originally began in 1950s using superfluid
$^{4}$He, and much theoretical, numerical, and experimental effort has been
devoted to the field. Superfluid $^{3}$He, discovered in 1972, presented a
system with a variety of quantized vortices characteristic of p-wave
superfluids.} is a vortex of inviscid superflow, and any rotational motion
of a superfluid is sustained by quantized vortices. A quantized vortex is
stable and well-defined topological defect, very different from classical
vortices in a conventional fluid. Hydrodynamics dominated by quantized
vortices is called \textit{quantum superfluid dynamics}, and turbulence
comprised of quantized vortices is known as \textit{quantum turbulence} (QT)
\cite{TsVor1,TsVor2}.

Liquid $^{4}$He enters a superfluid state below the $\lambda $ point ($%
T_{\lambda }=$ 2.17 K) with Bose--Einstein condensation of the $^{4}$He
atoms.\footnote{%
The characteristic phenomena of superfluidity were experimentally discovered
in the 1930s by L. Kapitza \cite{Kapitza} and Allen \textit{et al.}\cite%
{Allen}. The hydrodynamics of superfluid helium are well described by the
two-fluid model proposed by Landau \cite{Landau} and Tisza \cite{Tisza}.
According to the two-fluid model, the system consists of an inviscid
superfluid (density $\rho _{\mathrm{x}}$) and a viscous normal fluid
(density $\rho _{\mathrm{n}}$) with two independent velocity fields $\mathbf{%
v}_{\mathrm{x}}$ and $\mathbf{v}_{\mathrm{n}}$. The mixing ratio of the two
fluids depends on the temperature. As the temperature is reduced below the $%
\lambda $ point, the ratio of the superfluid component increases, and the
fluid becomes entirely superfluid below about 1 K. The two-fluid model
successfully explained the phenomena of superfluidity, while it was known in
1940s that superfluidity breaks down when it flows fast \cite{GM} and this
phenomenon was not explained through the two-fluid model. This was later
found to be caused by turbulence of the superfluid component due to random
motion of quantized vortices.} The $\lambda $ transition is closely related
to the Bose-Einstein condensation of $^{4}$He atoms, as first proposed by
\cite{London}. The Bose-condensed system exhibits the macroscopic
wave-function $\psi (\mathbf{x},t)=|\psi (\mathbf{x},t)|\mathrm{e}^{\mathrm{i%
}\theta (\mathbf{x},t)}$ as an order parameter. The superfluid velocity
field is given by $\mathbf{v}_{\mathrm{x}}=(\hbar /m)\nabla \theta $, with
boson mass $m$, representing the potential flow. Since the macroscopic wave
function should be single-valued for the space coordinate $\mathbf{x}$, the
circulation $\Gamma =\oint \mathbf{v}\cdot d\mathbf{\ell }$ for an arbitrary
closed loop in the fluid is quantized by the quantum $\kappa =h/m$. A vortex
with quantized circulation is called a quantized vortex. Any rotational
motion of a superfluid is sustained only by quantized vortices.\footnote{%
A quantized vortex is a topological defect characteristic of a
Bose--Einstein condensate, and is different from a vortex in a classical
viscous fluid. First, the circulation is quantized, which is contrary to a
classical vortex that can have any circulation value. Second, a quantized
vortex is a vortex of inviscid superflow. Thus, it cannot decay by the
viscous diffusion of vorticity that occurs in a classical fluid. Third, the
core of a quantized vortex is very thin, of the order of the coherence
length, which is only a few angstroms in superfluid $^{4}$He. Because the
vortex core is very thin and does not decay by diffusion, it is always
possible to identify the position of a quantized vortex in the fluid. These
properties make a quantized vortex more stable and definite than a classical
vortex.}

The idea of quantized circulation was first proposed by L. Onsager, for a
series of annular rings in a rotating superfluid \cite{Onsager}. R. Feynman
considered that a vortex in a superfluid could take the form of a vortex
filament, with the quantized circulation $\kappa $ and a core of atomic
dimension \cite{Feynman}.\footnote{%
Early experimental studies on superfluid hydrodynamics focused primarily on
thermal counterflow. The flow is driven by an injected heat current, and the
normal fluid and superfluid flow in opposite directions. The superflow was
found to become dissipative when the relative velocity between the two
fluids exceeds a critical value \cite{GM}. Gorter and Mellink attributed the
dissipation to mutual friction between two fluids, and considered the
possibility of superfluid turbulence. Feynman proposed a turbulent
superfluid state consisting of a tangle of quantized vortices \cite{Feynman}%
. Hall and Vinen performed the experiments of second sound attenuation in
rotating $^{4}$He, and found that the mutual friction arises from
interaction between the normal fluid and quantized vortices \cite%
{HallVinen56a,HallVinen56b}; second sound refers to entropy wave in which
superfluid and normal fluid oscillate oppositely, and its propagation and
attenuation give the information of the vortex density in the fluid.} Vinen
confirmed Feynman's findings experimentally, by showing that the dissipation
arises from mutual friction between vortices and the normal flow \cite%
{Vinen57a,Vinen57b,Vinen57c,Vinen57d}. Vinen also succeeded in observing
quantized circulation using vibrating wires in rotating superfluid $^{4}$He
\cite{Vinen61}. Subsequently, many experimental studies have examined
\textit{superfluid turbulence} (ST) in thermal counterflow systems, and have
revealed a variety of physical phenomena \cite{Tough82}. Since the dynamics
of quantized vortices are nonlinear and non-local, it has not been easy to
quantitatively understand these observations on the basis of vortex
dynamics. Schwarz clarified the picture of ST based on tangled vortices by
numerical simulation of the quantized vortex filament model in the thermal
counterflow \cite{Schwarz85,Schwarz88}. However, since the thermal
counterflow has no analogy in conventional fluid dynamics, this study was
not helpful in clarifying the relationship between ST and \textit{classical
turbulence} (CT). Superfluid turbulence is often called quantum turbulence
(QT), which emphasizes the belief that it is comprised of quantized vortices
\cite{TsVor1,TsVor2}.

\subsection{Bose-Einstein Crowd Superfluids}

Recall from the first section, what happens if we rotate a cylindrical
vessel with a classical viscous fluid inside. Even if the fluid is initially
at rest, it starts to rotate and eventually reaches a steady rotation with
the same rotational speed as the vessel. In that case, one can say that the
system contains a vortex that mimics solid-body rotation. A rotation of
arbitrary angular velocity can be sustained by a single vortex.

However, this does not occur in a \emph{quantum superfluid}. Because of
quantization of circulation, \emph{superfluids} respond to rotation, not
with a single vortex, but with a \emph{lattice of quantized vortices}. R.
Feynman noted that in uniform rotation with angular velocity $\Omega $ the
curl of the superfluid velocity is the circulation per unit area, and since
the curl is $2\Omega $, a lattice of quantized vortices with number density $%
n_{0}=\text{curl }v_{x}/\kappa =2\Omega /\kappa $ (the `Feynman rule')
arranges itself parallel to the rotation axis \cite{Feynman}. Such
experiments were performed for superfluid $^{4}$He: Packard \textit{et al.}
visualized vortex lattices on the rotational axis by trapping electrons
along the cores \cite{Williams74,Yarmchuck82}. This idea has also been
applied to atomic Bose-Einstein condensates. Several groups have observed
vortex lattices in rotating BECs\footnote{%
Among them, Madison \textit{et al}. directly observed nonlinear processes
such as vortex nucleation and lattice formation in a rotating $^{87}$Rb BEC
\cite{Madison01}. By sudden application of a rotation along the trapping
potential, an initially axi-symmetric condensate undergoes a collective
quadrupole oscillation to an elliptically deformed condensate. This
oscillation continues for a few hundred milliseconds with gradually
decreasing amplitude. Then the axial symmetry of the condensate is recovered
and vortices enter the condensate through its surface, eventually settling
into a lattice configuration.} \cite{Abo01, Madison00, Madison01,Matthews99}.

This observation has been reproduced by a simulation of the Gross-Pitaevskii
(GP) equation for the \emph{macroscopic wave-function} $\psi (\mathbf{x}%
,t)=|\psi (\mathbf{x},t)|\mathrm{e}^{\mathrm{i}\theta (\mathbf{x},t)}$ in 2D
\cite{Tsubota02,Kasamatsu03} and 3D \cite{Kasamatsu05a} spaces. In a weakly
interacting Bose system, the macroscopic wave-function $\psi (\mathbf{x},t)$
appears as the order parameter of the Bose--Einstein condensate
(representing our quantum crowd superfluid model) obeying the GP equation,
or the modified cubic NLS equation (extended by the linear term $-\mu \psi $%
):
\begin{equation}
\mathrm{i}\hbar \partial _{t}\psi =-\frac{\hbar ^{2}}{2m}\Delta \psi +V|\psi
|^{2}\psi -\mu \psi .  \label{gpeq}
\end{equation}%
Writing $\psi =|\psi |\exp (\mathrm{i}\theta )$, the squared amplitude $%
|\psi |^{2}$ is the \emph{crowd superfluid density} and the gradient of the
phase $\theta $ gives the \emph{crowd superfluid velocity} $\mathbf{v}%
_{x}=(\hbar /m)\nabla \theta $, corresponding to frictionless flow of the
crowd. This relation causes quantized vortices to appear with quantized
crowd circulation. The only characteristic scale of the GP model is the
coherence length defined by $\xi =\hbar /(\sqrt{2mV}\,|\psi |)$, which
determines the crowd vortex core size. The GP model can explain not only the
crowd vortex dynamics but also phenomena related to vortex cores, such as
crowd reconnection and nucleation.

\begin{figure}[h]
\centerline{\includegraphics[width=14cm]{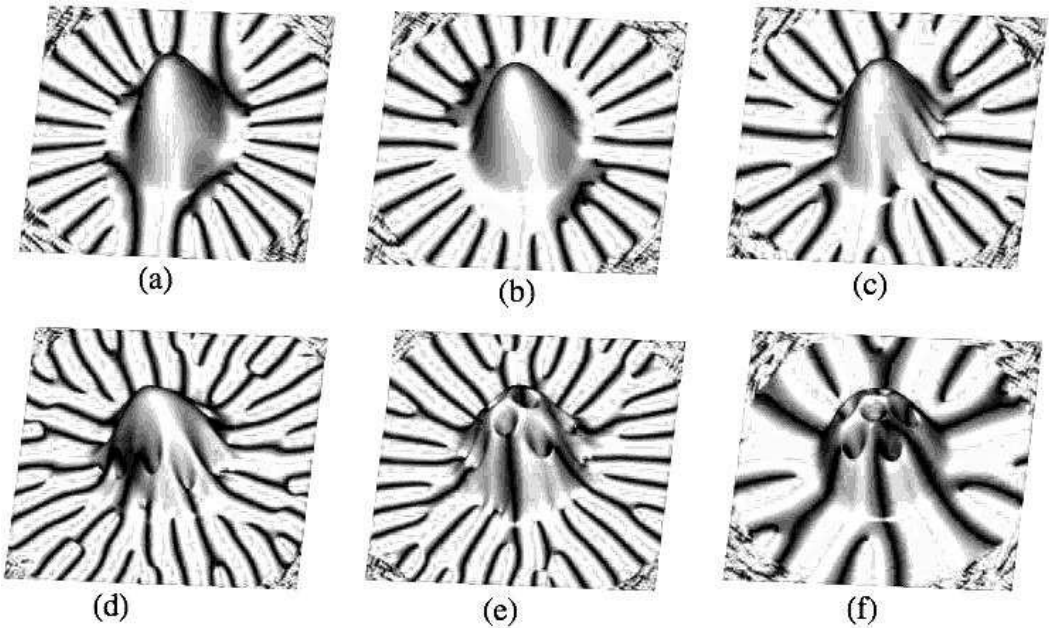}}
\caption{A typical crowd vortex lattice formation (modified and adapted from
\protect\cite{TsVor1}).}
\label{VortLat1}
\end{figure}
A typical 2D numerical simulation of equation (\ref{gpeq}) (adapted from
\cite{Tsubota02,Kasamatsu03}) for the crowd vortex lattice formation is
shown in Figure \ref{VortLat1}, where the crowd superfluid density and the
phase are displayed together. The trapping potential is
\begin{equation*}
V_{\mathrm{ex}}=\frac{1}{2}m\omega ^{2}[(1+\epsilon _{x})x^{2}+(1+\epsilon
_{y})y^{2}],
\end{equation*}%
where $\omega =2\pi \times 219$ Hz, and the parameters $\epsilon _{x}$ and $%
\epsilon _{y}$ describe small deviations from axisymmetry corresponding to
experiments \cite{Madison00, Madison01}. Following \cite{TsVor1,TsVor2}, we
first prepare an equilibrium condensate trapped in a stationary potential;
the size of the condensate cloud is determined by the Thomas-Fermi radius $%
R_{\mathrm{TF}}$. When we apply a rotation with $\Omega =0.7\omega $, the
condensate becomes elliptic and performs a quadrupole oscillation [Fig. \ref%
{VortLat1}(a)]. Then, the boundary surface of the condensate becomes
unstable and generates ripples that propagate along the surface [Fig. \ref%
{VortLat1}(b)]. As stated previously, it is possible to identify quantized
vortices in the phase profile also. As soon as the rotation starts, many
vortices appear in the low-density region outside of the condensate [Fig. %
\ref{VortLat1}(a)]. Since quantized vortices are excitations, their
nucleation increases the energy of the system. Because of the low density in
the outskirts of the condensate, however, their nucleation contributes
little to the energy and angular momentum.\footnote{%
Actually the vortex-antivortex pairs are nucleated in the low-density
region. Then the vortices parallel to the rotation are dragged into the
Thomas-Fermi surface, while the antivortices are repelled to the outskirts.}
Since these vortices outside of the condensate are not observed in the
density profile, they are called `ghost vortices'. Their movement toward the
Thomas-Fermi surface excites ripples [Fig. \ref{VortLat1}(b)]. It is not
easy for these ghost vortices to enter the condensate, because that would
increase both the energy and angular momentum. Only some vortices enter the
condensate cloud to become `real vortices' wearing the usual density profile
of quantized vortices [Fig. \ref{VortLat1}(d)], eventually forming a vortex
lattice [Fig. \ref{VortLat1}(e) and (f)]. The number of vortices forming a
lattice is given by `Feynman's rule' $n_{0}=2\Omega /\kappa $. The numerical
results agree quantitatively with these observations. Here we remark on the
essence of nonlinear dynamics. The initial state has no vortices in the
absence of rotation. The final state is a vortex lattice corresponding to
rotational frequency $\Omega $. In order to go from the initial to the final
state, the system makes use of as many excitations as possible, such as
vortices, quadruple oscillation, and surface waves. These experimental and
theoretical results demonstrate typical behavior of quantum fluid dynamics
in atomic BECs \cite{TsVor1,TsVor2}.

\subsection{Kolmogorov Energy Spectra}

The Kolmogorov energy spectra were confirmed for both decaying \cite%
{Kobayashi05a} and steady \cite{Kobayashi05b} QT by the GP model. The
normalized GP equation is:
\begin{equation}
\mathrm{i}\partial _{t}\psi =-\frac{1}{2}\Delta \psi +V|\psi |^{2}\psi -\mu
\psi ,  \label{eq-GP}
\end{equation}%
which determines the dynamics of the macroscopic wave-function $\psi (%
\mathbf{x},t)=f(\mathbf{x},t)\exp [\mathrm{i}\phi (\mathbf{x},t)]$. The
crowd superfluid dynamics in the GP model are compressible. The total number
of crowd agents is $N=\int |\psi |^{2}d\mathbf{x}$ and the total crowd
energy is:
\begin{equation*}
E(t)=\frac{1}{N}\int^{\ast }\psi \left( -\Delta +\frac{V}{2}f^{2}\right)
\psi \,d\mathbf{x},
\end{equation*}%
as represented by the sum of the interaction energy $E_{int}(t)$, the
quantum energy $E_{q}(t)$, and the kinetic energy $E_{kin}(t)$ \cite%
{Nore97a, Nore97b},
\begin{equation*}
E_{int}(t)=\frac{V}{2N}\int \,f^{4}d\mathbf{x},\qquad E_{q}(t)=\frac{1}{N}%
\int [\nabla f]^{2}d\mathbf{x},\qquad E_{kin}(t)=\frac{1}{N}\int [f\nabla
\phi ]^{2}d\mathbf{x}.
\end{equation*}%
The kinetic energy is further divided into a compressible part $%
E_{kin}^{c}(t)$ due to compressible excitations and an incompressible part $%
E_{kin}^{i}(t)$ due to vortices. The Kolmogorov spectrum is expected for $%
E_{kin}^{i}(t)\footnote{%
The failure to obtain a Kolmogorov law in the pure GP model \cite{Nore97a,
Nore97b} is attributable to the following (see \cite{TsVor1,TsVor2}). The
simulations showed that $E_{kin}^{i}(t)$ decreases and $E_{kin}^{c}(t)$
increases while the total energy $E(t)$ is conserved because many
compressible excitations are created through vortex reconnections \cite%
{Leadbeater01, Ogawa02} and disturb the Richardson cascade of quantized
vortices. Kobayashi \emph{et al.} overcame these difficulties and obtained a
Kolmogorov spectrum in QT that revealed an energy cascade \cite%
{Kobayashi05a, Kobayashi05b}. By performing numerical calculations of the
Fourier-transformed GP equation with dissipation, they confirmed the
Kolmogorov spectra for decaying turbulence \cite{Kobayashi05a}. To obtain a
turbulent state, they started the calculation from an initial configuration
in which the density was uniform and the phase of the wave-function had a
random spatial distribution. The initial state was dynamically unstable and
soon developed turbulence with many vortex loops. The spectrum $%
E_{kin}^{i}(k,t)$ was then found to obey the Kolmogorov law.}$ \cite%
{TsVor1,TsVor2}.

\section{A Variety of Crowd Waves}

\subsection{Crowd Shock-Waves, Solitons and Rogue Waves}

The general crowd NLS equation (\ref{nlsGen}) was exactly solved in \cite%
{LiuEtAl01,LiuFan05,IvCogComp}\ using the power series expansion method of {%
Jacobi elliptic functions} \cite{AbrSte}. Consider the $\psi -$function
describing a single plane wave, with the wave number $k$ and circular
frequency $\omega $:
\begin{equation}
\psi (x,t)=\phi (\xi )\,\mathrm{e}^{\mathrm{i}(kx-\omega t)},\qquad \text{%
with \ }\xi =x-kt\text{ \ and \ }\phi (\xi )\in \mathbb{R}.  \label{subGen}
\end{equation}%
Its substitution into the NLS equation (\ref{nlsGen}) gives the nonlinear
crowd oscillator ODE:
\begin{equation}
\phi ^{\prime \prime }(\xi )+[\omega -\frac{1}{2}k^{2}]\,\phi (\xi )-V\phi
^{3}(\xi )=0.  \label{15e}
\end{equation}

\bigskip We can seek a solution $\phi (\xi )$ for (\ref{15e}) as a linear
function \cite{LiuFan05}
\begin{equation*}
\phi (\xi )=a_{0}+a_{1}\mathrm{sn}(\xi ),
\end{equation*}%
where $\mathrm{sn}(x)=\mathrm{sn}(x,m)$ are Jacobi elliptic sine functions
with \textit{elliptic modulus} $m\in \lbrack 0,1]$, such that $\mathrm{sn}%
(x,0)=\sin (x)\ $and $\mathrm{sn}(x,1)=\mathrm{\tanh }(x)$. The solution of (%
\ref{15e}) was calculated in \cite{IvCogComp} to be
\begin{eqnarray*}
\phi (\xi ) &=&\pm m\sqrt{\frac{1}{V}}\,\mathrm{sn}(\xi ),\qquad ~\text{for~~%
}m\in \lbrack 0,1];~~\text{and} \\
\phi (\xi ) &=&\pm \sqrt{\frac{1}{V}}\,\mathrm{\tanh }(\xi ),\qquad \text{%
for~~}m=1.
\end{eqnarray*}%
This gives the exact periodic solution of (\ref{nlsGen}) as \cite{IvCogComp}
\begin{eqnarray}
\psi _{1}(x,t) &=&\pm m\sqrt{\frac{1}{V(w)}}\,\mathrm{sn}(x-kt)\,\mathrm{e}^{%
\mathrm{i}[kx-\frac{1}{2}t(1+m^{2}+k^{2})]},\qquad \text{for~~}m\in \lbrack
0,1);  \label{sn1} \\
\psi _{2}(x,t) &=&\pm \sqrt{\frac{1}{V(w)}}\,\mathrm{\tanh }(x-kt)\,\mathrm{e%
}^{\mathrm{i}[kx-\frac{1}{2}t(2+k^{2})]},\qquad \ \ \ \ \text{for~~}m=1,
\label{tanh1}
\end{eqnarray}%
where (\ref{sn1}) defines the general solution, while (\ref{tanh1}) defines
the \emph{crowd envelope shock-wave}\footnote{%
A shock wave is a type of fast-propagating nonlinear disturbance that
carries energy and can propagate through a medium (or, field). It is
characterized by an abrupt, nearly discontinuous change in the
characteristics of the medium. The energy of a shock wave dissipates
relatively quickly with distance and its entropy increases. On the other
hand, a soliton is a self-reinforcing nonlinear solitary wave packet that
maintains its shape while it travels at constant speed. It is caused by a
cancelation of nonlinear and dispersive effects in the medium (or, field).}
(or, `dark soliton') solution of the crowd NLS equation (\ref{nlsGen}).

Alternatively, if we seek a solution $\phi (\xi )$ as a linear function of
Jacobi elliptic cosine functions, such that $\mathrm{cn}(x,0)=\cos (x)$ and $%
\mathrm{cn}(x,1)=$\textrm{$\ $}$\mathrm{sech}(x)$,\footnote{%
A closely related solution of an anharmonic oscillator ODE:
\begin{equation*}
\phi ^{\prime \prime }(s)+\phi (s)+\phi ^{3}(s)=0
\end{equation*}%
is given by
\begin{equation*}
\phi (s)=\sqrt{\frac{2m}{1-2m}}\,\text{cn}\left( \sqrt{1+\frac{2m}{1-2m}}%
~s,\,m\right) .
\end{equation*}%
}
\begin{equation*}
\phi (\xi )=a_{0}+a_{1}\mathrm{cn}(\xi ),
\end{equation*}%
then we get \cite{IvCogComp}
\begin{eqnarray}
\psi _{3}(x,t) &=&\mp m\sqrt{\frac{1}{V(w)}}\,\mathrm{cn}(x-kt)\,\mathrm{e}^{%
\mathrm{i}[kx-\frac{1}{2}t(1-2m^{2}+k^{2})]},\qquad \text{for~~}m\in \lbrack
0,1);  \label{cn1} \\
\psi _{4}(x,t) &=&\mp \sqrt{\frac{1}{V(w)}}\,\mathrm{sech}(x-kt)\,\mathrm{e}%
^{\mathrm{i}[kx-\frac{1}{2}t(k^{2}-1)]},\qquad \qquad \text{for~~}m=1,
\label{sech1}
\end{eqnarray}%
where (\ref{cn1}) defines the general solution, while (\ref{sech1}) defines
the \emph{crowd envelope solitary-wave} (or, `bright soliton') solution of
the crowd NLS equation (\ref{nlsGen}).

In all four solution expressions (\ref{sn1}), (\ref{tanh1}), (\ref{cn1}) and
(\ref{sech1}), the adaptive crowd potential $V(w)$ is yet to be calculated
using either unsupervised Hebbian learning, or supervised
Levenberg--Marquardt algorithm (see, e.g. \cite{NeuFuz}). In this way, the
NLS equation (\ref{nlsGen}) becomes a \emph{quantum neural network }\cite%
{QnnBk}. Any kind of numerical analysis can be easily performed using above
closed-form crowd solutions $\psi _{i}(x,t)~~(i=1,...,4)$ as initial
conditions.

In addition, two new wave-solutions of the crowd NLS equation (\ref{nlsGen})
have been recently provided in \cite{Yan}, in the form of rogue waves,%
\footnote{%
Rogue waves are also known as \textit{freak waves}, \textit{monster waves},
\textit{killer waves}, \textit{giant waves} and \textit{extreme waves}. They
are found in various media, including optical fibers \cite{Solli}. The basic
rogue wave solution was first presented by Peregrine~\cite{Peregrine} to
describe the phenomenon known as \emph{Peregrine soliton} (or Peregrine
breather).} using the deformed Darboux transformation method developed in
\cite{Akhmediev}:

\begin{enumerate}
\item The \emph{one-rogon} crowd solution: {\small
\begin{equation}
\psi _{1\mathrm{rogon}}(x,t)=\alpha \sqrt{\frac{1}{2V}}\left[ 1-\frac{%
4(1+\alpha ^{2}t)}{1+2\alpha ^{2}(x-kt)^{2}+\sigma ^{2}\alpha ^{4}t^{2}}%
\right] \,\mathrm{e}^{\mathrm{i}[kx+1/2(\alpha ^{2}-k^{2})t]},\quad V>0,
\label{rog1}
\end{equation}%
} where $\alpha $ and $k$ denote the crowd scaling and gauge.

\item The \emph{two-rogon} crowd solution:
\begin{equation}
\psi _{2\mathrm{rogon}}(x,t)=\alpha \sqrt{\frac{1}{2V}}\left[ 1+\frac{%
P_{2}(x,t)+\mathrm{i}Q_{2}(x,t)}{R_{2}(x,t)}\right] \,\mathrm{e}^{\mathrm{i}%
[kx+1/2(\alpha ^{2}-k^{2})t]},\quad V>0,  \label{rog2}
\end{equation}%
where $P_{2},Q_{2},R_{2}$ are certain polynomial functions of $x$ and $t$.
\end{enumerate}

Both rogon crowd solutions can be easily made adaptive by introducing a set
of `synaptic weights' for nonlinear data fitting, in the same way as before.

\subsection{Collision of Two Crowds}

Next, a bidirectional quantum neural network resembling the strong crowd
coupling model (\ref{strongNLS}) has been formulated in \cite{IvCogComp} as
a self-organized system of two coupled NLS equations:
\begin{eqnarray}
\text{Red NLS :}\quad \mathrm{i}\partial _{t}\sigma &=&-\frac{1}{2}\partial
_{xx}\mathcal{\sigma }+V(w)\left( |\mathcal{\sigma }|^{2}+|\psi |^{2}\right)
\mathcal{\sigma },  \label{stochVol} \\
\text{Blue NLS :}\quad \mathrm{i}\partial _{t}\psi &=&-\frac{1}{2}\partial
_{xx}\psi +V(w)\left( |\mathcal{\sigma }|^{2}+|\psi |^{2}\right) \psi .
\label{stochPrice}
\end{eqnarray}%
In this coupled model, the $\sigma $--NLS (\ref{stochVol}) governs the $%
(x,t)-$evolution of the red crowd, which plays the role of a nonlinear
coefficient in the blue crowd (\ref{stochPrice}); the $\psi $--NLS (\ref%
{stochPrice}) defines the $(x,t)-$evolution of the blue crowd, which plays
the role of a nonlinear coefficient in the red crowd (\ref{stochVol}). The
purpose of this coupling is to generate the \textit{crowd leverage effect}
(similar to stock leverage effect in which stock volatility is (negatively)
correlated to stock returns. This bidirectional associative memory
effectively performs quantum neural computation \cite{QnnBk}, by giving a
spatiotemporal and quantum generalization of Kosko's BAM family of neural
networks \cite{Kosko1,Kosko2}. In addition, the shock-wave and solitary-wave
nature of the coupled NLS equations may describe brain-like effects
frequently occurring in crowd dynamics: propagation, reflection and
collision of shock and solitary waves (see \cite{Han}).

\begin{figure}[tbh]
\centering \includegraphics[width=14cm]{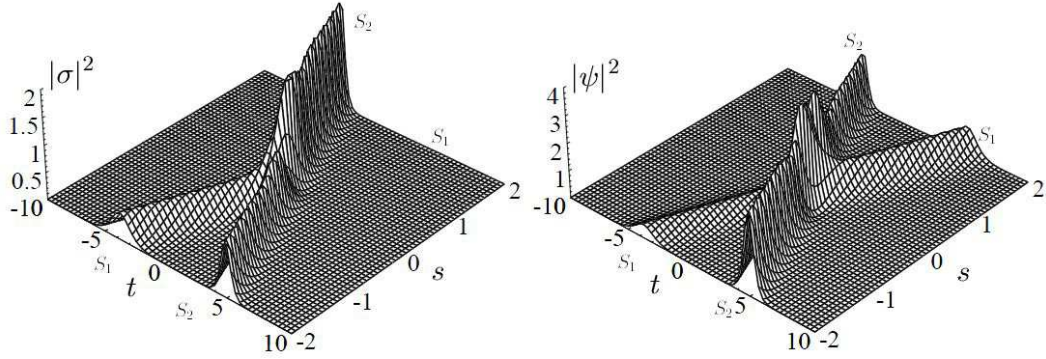}
\caption{Hypothetical crowd--collision scenario of the Manakov 2--soliton (%
\protect\ref{ManSol}). Due to symmetry of the Manakov system, the two crowds ($\psi$ and $\sigma$) can exchange their roles.}
\label{SolitonCollision}
\end{figure}

The coupled crowd NLS-system (\ref{stochVol})--(\ref{stochPrice}), without
an embedded $w-$learning (i.e., for constant $V$), actually defines the
well-known \emph{Manakov system},\footnote{%
Manakov system has been used to describe the interaction between wave
packets in dispersive conservative media, and also the interaction between
orthogonally polarized components in nonlinear optical fibres (see, e.g.
\cite{Kerr,Yang1} and references therein).} proven by S. Manakov in 1973
\cite{manak74} to be completely integrable, by the existence of infinite
number of involutive integrals of motion.Manakov's own method was based on
the \textit{Lax pair representation}.\footnote{%
The Manakov system (\ref{stochVol})--(\ref{stochPrice}) has the following
Lax pair \cite{Lax} representation:
\begin{eqnarray}
\partial _{x}\phi =M\phi \ \text{\ \ \ and \ \ }\partial _{t}\phi =B\phi ,%
\text{ \ \ or \ \ }\partial _{x}B-\partial _{t}M=[M,B],\qquad \text{with}
\label{man2} \\
M(\lambda )=\left(
\begin{array}{ccc}
-\mathrm{i}\lambda & \psi _{1} & \psi _{2} \\
-\psi _{1} & \mathrm{i}\lambda & 0 \\
-\psi _{2} & 0 & \mathrm{i}\lambda%
\end{array}
\right) \qquad \text{and}  \notag \\
B(\lambda )=-\mathrm{i}\left(
\begin{array}{ccc}
2\lambda ^{2}-|\psi _{1}|^{2}-|\psi _{2}|^{2} & 2\mathrm{i}\psi _{1}\lambda
-\partial _{x}\psi _{1} & 2\mathrm{i}\psi _{2}\lambda -\partial _{x}\psi _{2}
\\
-2\mathrm{i}\psi _{1}^{\ast }\lambda -\partial _{x}\psi _{1}^{\ast } &
-2\lambda ^{2}+|\psi _{1}|^{2} & \psi _{1}^{\ast }\psi _{2} \\
-2\mathrm{i}\psi _{2}^{\ast }\lambda -\partial _{x}\psi _{2}^{\ast } & \psi
_{1}\psi _{2}^{\ast } & -2\lambda ^{2}+|\psi _{2}|^{2}%
\end{array}
\right).  \notag
\end{eqnarray}%
} It admits both `bright' and `dark' soliton solutions. The simplest
solution of (\ref{stochVol})--(\ref{stochPrice}), the so-called \textit{%
Manakov bright 2--soliton} (see Figure \ref{SolitonCollision}), has the form
resembling that of the sech-solution (\ref{sech1}) (see \cite%
{Benney,Zakharov,Hasegawa,Radhakrishnan,Agrawal,Yang,Elgin}), and is
formally defined by:
\begin{equation}
\mathbf{\psi }_{\mathrm{xol}}(x,t)=2b\,\mathbf{c\,}\mathrm{sech}[2b(x+4at)]\,%
\mathrm{e}^{-2\mathrm{i}(2a^{2}t+ax-2b^{2}t)},  \label{ManSol}
\end{equation}%
where $\mathbf{\psi }_{\mathrm{xol}}(x,t)=\left(
\begin{array}{c}
\sigma (x,t) \\
\psi (x,t)%
\end{array}%
\right) $, $\mathbf{c}=(c_{1},c_{2})^{T}$ is a unit vector such that $%
|c_{1}|^{2}+|c_{2}|^{2}=1$. Real-valued parameters $a$ and $b$ are some
simple functions of $(V,k)$, which can be determined by the
Levenberg--Marquardt algorithm.

\subsection{Quantum Linear Crowd Waves}

In the case of very weak crowd heat potential $V(w)\ll 1$, we have $V(\psi
)\rightarrow 0,$ and therefore equation (\ref{nlsGen}) can be approximated
by a quantum-like \emph{crowd wave packet.} It is defined by a continuous
superposition of \emph{de Broglie's plane waves}, `physically' associated
with a free quantum particle of unit mass. This linear wave packet, given by
the time-dependent complex-valued wave function $\psi =\psi (x,t)$, is a
solution of the \emph{linear Schr\"{o}dinger equation} with zero potential
energy and the \textit{crowd Hamiltonian operator} $\hat{H}$. This equation
can be written as:
\begin{equation}
\mathrm{i}\partial _{t}\psi =\hat{H}\psi ,\qquad \text{where}\qquad \hat{H}=-%
\frac{1}{2}\partial _{xx}.  \label{sch1}
\end{equation}

Thus, we consider the $\psi -$function describing a single de Broglie's
plane wave, with the wave number $k$, linear momentum $p=k,$ wavelength $%
\lambda _{k}=2\pi /k,$\ angular frequency $\omega _{k}=k^{2}/2,$ and
oscillation period $T_{k}=2\pi /\omega _{k}=4\pi /k^{2}$. It is defined by
(compare with \cite{Griffiths,Thaller,QuLeap})
\begin{equation}
\psi _{k}(x,t)=A\mathrm{e}^{\mathrm{i}(kx-\omega _{k}t)}=A\mathrm{e}^{%
\mathrm{i}(kx-{\frac{k^{2}}{2}}t)}=A\cos (kx-{\frac{k^{2}}{2}}t)+A\mathrm{i}%
\sin (kx-{\frac{k^{2}}{2}}t),  \label{Broglie}
\end{equation}%
where $A$ is the amplitude of the wave, the angle $(kx-\omega _{k}t)=(kx-{%
\frac{k^{2}}{2}}t)$\ represents the phase of the wave $\psi _{k}$ with the
\textit{crowd phase velocity}\emph{:} $v_{k}=\omega _{k}/k=k/2.$

The space-time wave function $\psi (x,t)$ that satisfies the linear Schr\"{o}%
dinger equation (\ref{sch1}) can be decomposed (using Fourier's separation
of variables) into the spatial part $\phi (x)\,$\ and the temporal part $%
\mathrm{e}^{-\mathrm{i}\omega t}\ $as:
\begin{equation*}
\psi (x,t)=\phi (x)\,\mathrm{e}^{-\mathrm{i}\omega t}=\phi (x)\,\mathrm{e}^{-%
\mathrm{i}Et}=\phi (x)\,\mathrm{e}^{-\frac{\mathrm{i}}{2}k^{2}t},
\end{equation*}%
where Planck's \emph{energy quantum} of the \textit{crowd wave} $\psi _{k}$
is given by: $E_{k}=\omega _{k}=\frac{1}{2}k^{2}.$

The spatial part, representing \emph{stationary }(or,\emph{\ amplitude})%
\emph{\ wave function}, $\phi (x)=A\mathrm{e}^{\mathrm{i}kx},$ satisfies the
\emph{crowd harmonic oscillator,} which can be formulated in several
equivalent forms:
\begin{equation}
\phi ^{\prime \prime }+k^{2}\phi =0,\qquad \phi ^{\prime \prime }+\left(
\frac{\omega _{k}}{v_{k}}\right) ^{2}\phi =0,\qquad \phi ^{\prime \prime
}+2E_{k}\phi =0.  \label{stac}
\end{equation}

From the plane-wave expressions (\ref{Broglie}) we have: $\psi _{k}(x,t)=A%
\mathrm{e}^{\mathrm{i}(px-E_{k}t)}-$ for the wave going to the `right' and $%
\psi _{k}(x,t)=A\mathrm{e}^{-\mathrm{i}(px+E_{k}t)}-$ for the wave going to
the `left'.

The general solution to (\ref{sch1}) is formulated as a linear combination
of de Broglie's planar waves (\ref{Broglie}), comprising the crowd
wave-packet:
\begin{equation}
\psi (x,t)=\sum_{i=0}^{n}c_{i}\psi _{k_{i}}(x,t),\qquad (\text{with}\ n\in
\mathbb{N}).  \label{w-pack}
\end{equation}%
Its absolute square, $|\psi (x,t)|^{2},$ represents crowd's probability
density function at a time $t.$

The \textit{crowd group velocity} is given by: $\ v_{g}=d\omega _{k}/dk.$ It
is related to the crowd phase velocity $v_{k}$: $v_{g}=v_{k}-\lambda
_{k}dv_{k}/d\lambda _{k}.$ Closely related is the \emph{center} of the crowd
wave-packet (the point of maximum crowd amplitude), given by: $x=td\omega
_{k}/dk.$\newline

The following quantum-motivated assertions can be stated:

\begin{enumerate}
\item The total energy $E$ of an crowd wave-packet is (in the case of
similar plane waves) given by Planck's superposition of the energies $E_{k}$
of $n$ individual agents' waves: $E=n\omega _{k}=\frac{n}{2}k^{2},$ where $%
L=n$ denotes the \emph{angular momentum} of the crowd wave-packet,
representing the shift between its growth and decay, and \emph{vice versa.}

\item The average energy $\left\langle E\right\rangle $\ of an crowd
wave-packet is given by Boltzmann's partition function:
\begin{equation*}
\left\langle E\right\rangle =\frac{\sum_{n=0}^{\infty }nE_{k}\mathrm{e}^{-%
\frac{nE_{k}}{bT}}}{\sum_{n=0}^{\infty }\mathrm{e}^{-\frac{nE_{k}}{bT}}}=%
\frac{E_{k}}{\mathrm{e}^{\frac{E_{k}}{bT}}-1},
\end{equation*}%
where $b$ is the Boltzmann-like kinetic constant and $T$ is the crowd
`temperature'.

\item The energy form of the Schr\"{o}dinger equation (\ref{sch1}) reads: $%
E\psi =\mathrm{i}\partial _{t}\psi $.

\item The eigenvalue equation for the crowd Hamiltonian operator $\hat{H}$
is the \emph{stationary Schr\"{o}dinger equation:} $\ $%
\begin{equation*}
\hat{H}\phi (x)=E\phi (x),\qquad \text{or}\qquad E\phi (x)=-\frac{1}{2}%
\partial _{xx}\phi (x),
\end{equation*}%
which is just another form of the harmonic oscillator (\ref{stac}). It has
oscillatory solutions of the form:
\begin{equation*}
\phi _{E}(x)=c_{1}\mathrm{e}^{\mathrm{i}\sqrt{2E_{k}}\,x}+c_{2}\mathrm{e}^{-%
\mathrm{i}\sqrt{2E_{k}}\,x}\,,
\end{equation*}%
called \emph{energy eigen-states} with energies $E_{k}$ and denoted by: $%
\hat{H}\phi _{E}(x)=E_{k}\phi _{E}(x).$
\end{enumerate}

Now, given some initial crowd wave function, $\psi (x,0)=\psi _{0}(x),$ a
solution to the initial-value problem for the linear Schr\"{o}dinger
equation (\ref{sch1}) is, in terms of the pair of Fourier transforms $(%
\mathcal{F},\mathcal{F}^{-1}),$ given by (see \cite{Thaller})
\begin{equation}
\psi (x,t)=\mathcal{F}^{-1}\left[ \mathrm{e}^{-\mathrm{i}\omega t}\mathcal{F}%
(\psi _{0})\right] =\mathcal{F}^{-1}\left[ \mathrm{e}^{-\mathrm{i}{\frac{%
k^{2}}{2}}t}\mathcal{F}(\psi _{0})\right] .  \label{Fouri}
\end{equation}

For example (see \cite{Thaller}), suppose we have an initial crowd
wave-function at time $t=0$ given by the complex-valued Gaussian function:
\begin{equation*}
\psi (x,0)=\mathrm{e}^{-ax^{2}/2}\mathrm{e}^{\mathrm{i}kx},
\end{equation*}%
where $a$ is the width of the Gaussian, while $p$ is the average momentum of
the wave. Its Fourier transform, $\hat{\psi}_{0}(k)=\mathcal{F}[\psi (x,0)],$
is given by
\begin{equation*}
\hat{\psi}_{0}(k)=\frac{\mathrm{e}^{-\frac{(k-p)^{2}}{2a}}}{\sqrt{a}}.
\end{equation*}%
The solution at time $t$ of the initial value problem is given by
\begin{equation*}
\psi (x,t)=\frac{1}{\sqrt{2\pi a}}\int_{-\infty }^{+\infty }\mathrm{e}^{%
\mathrm{i}(kx-{\frac{k^{2}}{2}}t)}\,\mathrm{e}^{-\frac{a(k-p)^{2}}{2a}}\,dk,
\end{equation*}%
which, after some algebra becomes
\begin{equation*}
\psi (x,t)=\frac{\mathrm{\exp }(-\frac{ax^{2}-2\mathrm{i}xp+\mathrm{i}p^{2}t%
}{2(1+\mathrm{i}at)})}{\sqrt{1+\mathrm{i}at}},\qquad (\text{with \ }p=k).
\end{equation*}

As a simpler example,\footnote{%
An example of a more general Gaussian wave-packet solution of (\ref{sch1})
is given by:
\begin{equation*}
\psi (x,t)=\sqrt{\frac{\sqrt{a/\pi }}{1+\mathrm{i}at}}\,\exp \left( \frac{-%
\frac{1}{2}a(s-{s_{0}})^{2}-\frac{\mathrm{i}}{2}p_{0}^{2}t+\mathrm{i}p_{0}(s-%
{s_{0}})}{1+\mathrm{i}at}\right) ,
\end{equation*}%
where $s_{0},p_{0}$ are initial stock-price and average momentum, while $a$
is the width of the Gaussian. At time $t=0$ the `particle' is at rest around
$s=0$, its average momentum $p_{0}=0$. The wave function spreads with time
while its maximum decreases and stays put at the origin. At time $-t$ the
wave packet is the complex-conjugate of the wave-packet at time $t$.} if we
have an initial crowd wave-function given by the real-valued Gaussian
function,
\begin{equation*}
\psi (x,0)=\frac{\mathrm{e}^{-x^{2}/2}}{\sqrt[4]{\pi }},
\end{equation*}%
the solution of (\ref{sch1}) is given by the complex-valued $\psi -$%
function,
\begin{equation*}
\psi (x,t)=\frac{\mathrm{\exp }(-\frac{x^{2}}{2(1+\mathrm{i}t)})}{\sqrt[4]{%
\pi }\sqrt{1+\mathrm{i}t}}.
\end{equation*}

From (\ref{Fouri}) it follows that a stationary crowd wave-packet is given
by:
\begin{equation*}
\phi (x)=\frac{1}{\sqrt{2\pi }}\int_{-\infty }^{+\infty }\mathrm{e}^{\mathrm{%
i}kx}\,\hat{\psi}(k)\,dk,\qquad \text{where}\qquad \hat{\psi}(k)=\mathcal{F}%
[\phi (x)].
\end{equation*}%
As $|\phi (x)|^{2}$ is the stationary crowd PDF, we can calculate the
\textit{crowd expectation values} and the wave number of the whole crowd
wave-packet, consisting of $n$ measured plane waves, as:
\begin{equation}
\left\langle x\right\rangle =\int_{-\infty }^{+\infty }x|\phi
(x)|^{2}dx\qquad \text{and}\qquad \left\langle k\right\rangle =\int_{-\infty
}^{+\infty }k|\hat{\psi}(k)|^{2}dk.  \label{means}
\end{equation}%
The recordings of $n$ individual crowd plane waves (\ref{Broglie}) will be
scattered around the mean values (\ref{means}). The width of the
distribution of the recorded $x-$ and $k-$values are uncertainties $\Delta x$
and $\Delta k,$ respectively. They satisfy the Heisenberg-type uncertainty
relation:
\begin{equation*}
\Delta x\,\Delta k\geq \frac{n}{2},
\end{equation*}%
which imply the similar relation for the total crowd energy and time:
\begin{equation*}
\Delta E\,\Delta t\geq \frac{n}{2}.
\end{equation*}

\section{Conclusion}

In this paper we gave a formal mathematical and physical description of nonlinear phenomena in dynamics of human crowds. While Helbing discovered a phenomenon of crowd turbulence (see Introduction), we felt that equally important would be to model related but different crowd phenomena, such as solitons, rogue waves and shock waves. Our proposal, including both classical and quantum description of crowd turbulence, as well as both nonlinear and quantum crowd waves, provides a new basis for studying all these nonlinear phenomena in crowds.

\section{Appendix: Basic Lie Algebra Mechanics}

A \textit{manifold} $M$ is a topological space that on a small scale
(locally) resembles the Euclidean space. Manifolds are usually endowed with
a differentiable structure that allows one to do calculus and differential
equations, as well as a \textit{Riemannian metric} that allows one to
measure distances and angles. For example, Riemannian manifolds are the
configuration spaces for Lagrangian mechanics, while symplectic manifolds
are the phase spaces in the Hamiltonian mechanics. A \textit{diffeomorphism}
is an invertible function that maps one smooth (differentiable) manifold to
another, such that both the function and its inverse are smooth.

A \textit{Lie group} $G$ is a smooth manifold $M$ that has at the same time
a group $G-$structure consistent with its manifold $M-$structure in the
sense that \textit{group multiplication} ~$\mu :G\times G\rightarrow
G,~~(g,h)\mapsto gh$ and the \textit{group inversion} ~$\nu :G\rightarrow
G,~~g\mapsto g^{-1}$ are smooth maps. A point $e\in G$ is called the group
identity element.

A Lie group can \textit{act} on a smooth manifold $M$ by moving the points
of $M,$ denoted by $G\times M\rightarrow M.$Group action on a manifold
defines the \textit{orbit} of a point $m$ on a manifold $M,$ which is the
set of points on $M$ to which $m$ can be moved by the elements of a Lie
group $G$. The orbit of a point $m$ is denoted by $Gm=\{g\cdot m|g\in G\}.$

Let $G$ be a real Lie group. Its \textit{Lie algebra} $\mathfrak{g}$ is the
tangent space $TG_{e}$\ to the group $G$\ at the identity $e$ provided with
the \textit{Lie bracket} (\textit{commutator}) operation $[X,Y],$ which is
bilinear, skew-symmetric, and satisfies the \textit{Jacobi identity} (for
any three vector fields $X,Y,Z\in \mathfrak{g}$):
\begin{equation*}
\lbrack \lbrack X,Y],Z]=[X,[Y,Z]]-[X,[Y,Z]].
\end{equation*}%
Note that in Hamiltonian mechanics, Jacobi identity is satisfied by Poisson
brackets, while in quantum mechanics it is satisfied by operator commutators.

For example, $G=SO(3)$ is the group of rotations of 3D Euclidean space, i.e.
the configuration space of a rigid body fixed at a point. A motion of the
body is then described by a curve $g=g(t)$ in the group $SO(3)$. Its Lie
algebra $\mathfrak{g}=\mathfrak{so}(3)$\ is the 3D vector space of angular
velocities of all possible rotations. The commutator in this algebra is the
usual vector (cross) product.

A Lie group $G$ acts on itself by left and right translations: every element
$g\in G$ defines diffeomorphisms of the group onto itself (for every $h\in G$%
):
\begin{equation*}
L_{g}:G\rightarrow G,\qquad L_{g}h=gh;\qquad R_{g}:G\rightarrow G,\qquad
R_{g}h=hg.
\end{equation*}%
The induced maps of the tangent spaces are denoted by:

\begin{equation*}
L_{g\ast }:TG_{h}\rightarrow TG_{gh},\qquad R_{g\ast }:TG_{h}\rightarrow
TG_{hg}.
\end{equation*}

The diffeomorphism $R_{g^{-1}}L_{g}$ is an inner automorphism of the group $%
G $. It leaves the group identity $e$ fixed. Its derivative at the identity $%
e$ is a linear map from the Lie algebra $\mathfrak{g}$ to itself:%
\begin{equation*}
Ad_{g}:\mathfrak{g}\rightarrow \mathfrak{g},\qquad
Ad_{g}(R_{g^{-1}}L_{g})_{\ast e}
\end{equation*}%
is called the \textit{adjoint representation} of the Lie group $G$.

Referring to the previous example, a rotation velocity $\dot{g}$ of the
rigid body (fixed at a point) is a tangent vector to the Lie group $G=SO(3)$
at the point $g\in G$. To get the angular velocity, we must carry this
vector to the tangent space $TG_{e}$ of the group at the identity, i.e. to
its Lie algebra $\mathfrak{g}=\mathfrak{so}(3)$. This can be done in two
ways: by left and right translation, $L_{g}$ and $R_{g}$. As a result, we
obtain two different vector fields in the Lie algebra $\mathfrak{so}(3):$

\begin{equation*}
\omega _{c}=L_{g^{-1}\ast }\dot{g}\in \mathfrak{so}(3)\qquad \text{and\qquad
}\omega _{x}=R_{g^{-1}\ast }\dot{g}\in \mathfrak{so}(3),
\end{equation*}%
which are called the `angular velocity in the body' and the `angular
velocity in space,' respectively.

Now, left and right translations induce operators on the cotangent space $%
T^{\ast }G_{g}$\ dual to $L_{g\ast }$ and $R_{g\ast },$ denoted by (for
every $h\in G$):%
\begin{equation*}
L_{g}^{\ast }:T^{\ast }G_{gh}\rightarrow T^{\ast }G_{h},\qquad R_{g}^{\ast
}:T^{\ast }G_{hg}\rightarrow T^{\ast }G_{h}.
\end{equation*}%
The transpose operators $Ad_{g}^{\ast }:\mathfrak{g}\rightarrow \mathfrak{g}$
satisfy the relations $Ad_{gh}^{\ast }=Ad_{h}^{\ast }Ad_{g}^{\ast }$ (for
every $g,h\in G$) and constitute the \textit{co-adjoint representation} of
the Lie group $G$. The co-adjoint representation plays an important role in
all questions related to (left) invariant metrics on the Lie group.
According to A. Kirillov, the orbit of any vector field $X$ in a Lie algebra
$\mathfrak{g}$ in a co-adjoint representation $Ad_{g}^{\ast }$ is itself a
symplectic manifold and therefore a phase space for a Hamiltonian mechanical
system.

A Riemannian metric on a Lie group $G$ is called left-invariant if it is
preserved by all left translations $L_{g}$, i.e., if the derivative of left
translation carries every vector to a vector of the same length. Similarly,
a vector field $X$ on $G$ is called left--invariant if (for every $g\in G$) $%
L_{g}^{\ast }X=X$.

Again referring to the previous example of the rigid body, the dual space $%
\mathfrak{g}^{\ast }$ to the Lie algebra $\mathfrak{g}=\mathfrak{so}(3)$ is
the space of angular momenta $\mathbf{\pi }$. The kinetic energy $T$ of a
body is determined by the vector field of angular velocity in the body and
does not depend on the position of the body in space. Therefore, kinetic
energy gives a left-invariant Riemannian metric on the rotation group $%
G=SO(3)$.


\begin{thebibliography}{KK01}
\bibitem{NlDyn1} V. Ivancevic, D. Reid, E. Aidman, Crowd behavior dynamics:
entropic path-integral model. Nonl. Dyn. \textbf{59}, 351-373, (2010)

\bibitem{NlDyn2} V. Ivancevic, D. Reid, Crowd behavior dynamics: entropic
path-integral model. Nonl. Dyn. Entropic geometry of crowd dynamics. A
Chapter in Nonlear Dynamics (T. Evancs, Ed.), Intech, Vienna, (2010)

\bibitem{TopDual} V. Ivancevic, D. Reid, Geometrical and Topological Duality
in Crowd Dynamics. Int. J. Biomath. \textbf{3}(4), (2010), 493--507.

\bibitem{IvResonan} V. Ivancevic, D. Reid, Dynamics of Confined Crowds
Modelled using Entropic Stochastic Resonance and Quantum Neural Networks.
Int. J. Intel. Defence Sup. Sys. \textbf{2}(4), 269-289, (2009)

\bibitem{HelbingPRE1} Helbing, D., Molnar, P., Social force model for
pedestrian dynamics. Phys. Rev. E 1995, \textbf{51}(5), 4282--4286.

\bibitem{HelbingNature} Helbing, D., Farkas, I., Vicsek, T. Simulating
dynamical features of escape panic. Nature \textbf{407}, (2000), 487--490.

\bibitem{HelbingPRL} Helbing, D., Johansson, A., Mathiesen, J., Jensen,
M.H., Hansen, A. Analytical approach to continuous and intermittent
bottleneck flows. Phys. Rev. Lett.  \textbf{97}, (2006), 168001.

\bibitem{HelbingPRE} Helbing, D., Johansson, A., Zein Al-Abideen, H. The
Dynamics of Crowd Disasters: An Empirical Study. Phys. Rev. E \textbf{75},
(2007), 046109.

\bibitem{HelbingACS} Johansson, A., Helbing, D., Z. Al-Abideen, H.,
Al-Bosta, S. From Crowd Dynamics to Crowd Safety: A Video--Based Analysis.
Adv. Com. Sys. \textbf{11}(4), (2008), 497--527.

\bibitem{QnnBk} V. Ivancevic, T. Ivancevic, Quantum Neural Computation,
Springer, (2009)

\bibitem{NeuFuz} Ivancevic, V., Ivancevic, T., Neuro--Fuzzy Associative
Machinery for Comprehensive Brain and Cognition Modelling. Springer, Berlin,
(2007)

\bibitem{tao:cbms} Tao, T., Nonlinear dispersive equations: local and global
analysis, CBMS regional series in mathematics, (2006)

\bibitem{ArnoldMech} Arnold V.I., Mathematical Methods of Classical
Mechanics (2ed.), Springer, (1989)

\bibitem{ArnoldHidr} Arnold V.I., Khezin B., Topological Methods in
Hydrodynamics, Springer, (1998)

\bibitem{GaneshSprBig} Ivancevic, V., Ivancevic, T., Geometrical Dynamics of
Complex Systems: A Unified Modelling Approach to Physics Control
Biomechanics Neurodynamics and Psycho--Socio--Economical Dynamics. Springer:
Dordrecht, 2006.

\bibitem{GaneshADG} Ivancevic, V., Ivancevic, T., Applied Differential
Geometry: A Modern Introduction. World Scientific: Singapore, 2007.

\bibitem{Ruelle} Ruelle, D., Takens, F., On the nature of turbulence, Comm.
Math. Phys. \textbf{20}(2), (1971), 167-192; Comm. Math. Phys. \textbf{23}%
(3), (1971), 343-344.

\bibitem[KK01]{Kawahara} Kawahara, G., Kida, S.: Periodic motion embedded in
plane Couette turbulence: regeneration cycle and burst. J. Fluid Mech.
\textbf{449}, 291--300, (2001)

\bibitem{Kir} Kirchhoff, G.R., Vorlesungen \"{u}ber mathematische Physik.
Mechanik, Leipzig, Teubner, (1876), 466 pp.

\bibitem{TsVor1} Tsubota, M. Quantized vortices in superfluid helium and
Bose-Einstein condensates. J. Physics: Conf. Ser. \textbf{31}, 88--94, (2006)

\bibitem{TsVor2} Tsubota, M., Kasamatsu, K, Kobayashi, M. Quantized vortices
in superfluid helium and atomic Bose-Einstein condensates. arXiv:
cond-mat.quant-gas 1004.5458v2, (2010)

\bibitem{Pitaevskii} Pitaevskii, L. and Stringari, S. (2003). Bose-Einstein
Condensation. Oxford University Press, Oxford.

\bibitem{Kapitza} Kapitza, P., Viscosity of liquid helium below the $\lambda
$ point. Nature \textbf{141},  (1938), 74.

\bibitem{Allen} Allen, J.F., Misener, A.D., Flow of liquid helium II.
Nature, \textbf{141}, (1938), 75.

\bibitem{Landau} Landau, L. (1941). The theory of superfluidity of helium
II. J. Phys. U.S.S.R. \textbf{5}, 71-90.

\bibitem{Tisza} Tisza, L. (1938). Transport phenomena in helium II. Nature,
141, 913.

\bibitem{GM} Gorter, C J., Mellink, J.H. (1949). On the irreversible
processes in liquid helium II. Physica, \textbf{15}, 285-304.

\bibitem{London} London, F. (1938). On the Bose-Einstein condensation. Phys.
Rev. \textbf{54}, 947-954.

\bibitem{Onsager} Onsager, L. (1949). Nuovo Cimento Suppl. \textbf{6},
249-250.

\bibitem{Feynman} Feynman, R.P. (1955). Application of quantum mechanics to
liquid helium. Progress in Low Temperature Physics Vol.1(Gorter, C. J. ed.).
Amsterdam. North-Holland, 17-53.

\bibitem{HallVinen56a} Hall, H. E. and Vinen, W. F. (1956). The rotation of
liquid helium II I. Experiments on the propagation of second sound in
uniformly rotating helium II. Proc. Roy. Soc. London, A \textbf{238},
204-214.

\bibitem{HallVinen56b} Hall, H. E. and Vinen, W. F. (1956). The rotation of
liquid helium II II. The theory of mutual friction in uniformly rotating
helium II. Proc. Roy. Soc. London, A \textbf{238}, 215-234.

\bibitem{Vinen57a} Vinen, W.F. (1957). Mutual friction in a heat current in
liquid helium II I. Experiments on steady heat currents, Proc. Roy. Soc.
London, A \textbf{240}, 114-127.

\bibitem{Vinen57b} Vinen, W. F. (1957). Mutual friction in a heat current in
liquid helium II. II. Experiments on transient effects. Proc. Roy. Soc.
London, A \textbf{240}, 128-143.

\bibitem{Vinen57c} Vinen, W. F. (1957). Mutual friction in a heat current in
liquid helium II III. Theory of mutual friction. Proc. Roy. Soc. London, A
\textbf{242}, 493-515.

\bibitem{Vinen57d} Vinen, W. F. (1957). Mutual friction in a heat current in
liquid helium II IV. Critical heat currents in wide channels. Proc. Roy.
Soc. London, A \textbf{243}, 400-413.

\bibitem{Vinen61} Vinen, W. F. (1961). The detection of single quanta
circulation in liquid helium II. Proc. Roy. Soc. London, A \textbf{260},
218-236.

\bibitem{Tough82} Tough, J. T. (1982). Superfluid turbulence. Progress in
Low Temperature Physics Vol. 8 (Gorter, C. J. ed.). Amsterdam.
North-Holland, 133-220.

\bibitem{Schwarz85} Schwarz, K. W. (1985). Three-dimensional vortex dynamics
in superfluid $^{4}$He: Line-line and line-boundary interactions. Phys. Rev.
B \textbf{31}, 5782-5803.

\bibitem{Schwarz88} Schwarz, K. W. (1988). Three-dimensional vortex dynamics
in superfluid $^{4}$He: Homogeneous superfluid turbulence. Phys. Rev. B
\textbf{38}, 2398-2417.

\bibitem{Williams74} G.A. Williams and R.E. Packard, Photographs of
quantized vortex lines in rotating He II, Phys. Rev. Lett. \textbf{33}
(1974), 280--283.

\bibitem{Yarmchuck82} E.J. Yarmchuck and R.E. Packard, Photographic studies
of quantized vortex lines, J. Low Temp. Phys. 46 (1982), 479--515.

\bibitem{Matthews99} M.R. Matthews, B.P. Anderson, P. C. Haljan, D. S. Hall,
C. E. Wieman, and E. A. Cornell, Vortices in a Bose-Einstein condensate,
Phys. Rev. Lett. \textbf{83} (1999), 2498--2501.

\bibitem{Abo01} J.R. Abo-Shaeer, C. Raman, J. M. Vogels, and W. Ketterle,
Observation of vortex lattices in Bose-Einstein condensates, Science \textbf{%
292} (2001), 476--479.

\bibitem{Madison00} K.W. Madison, F. Chevy, W. Wlhlleben, and J. Dalibard,
Vortex formation in a stirred Bose-Einstein condensate, Phys. Rev. Lett.
\textbf{84} (2000), 806--809.

\bibitem{Madison01} K.W. Madison, F. Chevy, W. Wlhlleben, and J. Dalibard,
Statonary states of a rotating Bose-Einstein condensate: Routes to vortex
nucleation, Phys. Rev. Lett. \textbf{86} (2001), 4443--4446.

\bibitem{Tsubota02} M. Tsubota, K. Kasamatsu, and M. Ueda, Vortex lattice
formation in a rotating Bose-Einstein condensate, Phys. Rev. A \textbf{65}
(2002), 023603.

\bibitem{Kasamatsu03} K. Kasamatsu, M. Tsubota, and M. Ueda, Nonlinear
dynamics of vortex lattice formation in a rotating Bose-Einstein condensate,
Phys. Rev. A \textbf{67} (2003), 033610.

\bibitem{Kasamatsu05a} K. Kasamatsu, M. Machida, N. Sasa and M. Tsubota,
Three-dimensional dynamics of vortex-lattice formation in Bose-Einstein
condensate Phys. Rev. A \textbf{71} (2005), 063616.

\bibitem{Frisch} U. Frisch, Turbulence, Cambridge University Press,
Cambridge, 1995.

\bibitem{Kolmogorov41a} A.N. Kolmogorov, The local structure of turbulence
in incompressible viscous fluid for very large Reynolds number, Dokl. Akad.
Nauk SSSR 30 (1941), 299-303 [reprinted in Proc. Roy. Soc. A \textbf{434}
(1991), 9-13].

\bibitem{Kolmogorov41b} A.N. Kolmogorov, On degeneration (decay) of
isotropic turbulence in an incompressible viscous liquid, Dokl. Akad. Nauk
SSSR 31 (1941), 538-540 [reprinted in Proc. Roy. Soc. A \textbf{434} (1991),
15-17].

\bibitem{Leadbeater01} M. Leadbeater, T. Winiecki, D. C. Samuels, C. F.
Barenghi, and C. S. Adams, Sound emission due to superfluid vortex
reconnections, Phys. Rev. Lett. \textbf{86} (2001), 1410-1413.

\bibitem{Ogawa02} S. Ogawa, M. Tsubota, and Y. Hattori, Study of
reconnection and acoustic emission of quantized vortices in superfluid by
the numerical analysis of the Gross-Pitaevskii equation, J. Phys. Soc. Jpn.
\textbf{71} (2002), 813-821.

\bibitem{Kobayashi05a} M. Kobayashi and M. Tsubota, Kolmogorov spectrum of
superfluid turbulence: Numerical analysis of the Gross-Pitaevskii equation
with a small-scale dissipation, Phys. Rev. Lett. \textbf{94} (2005), 065302.

\bibitem{Kobayashi05b} M. Kobayashi and M. Tsubota, Kolmogorov spectrum of
quantum turbulence, J. Phys. Soc. Jpn. \textbf{74} (2005), 3248-3258.

\bibitem{Nore97a} C. Nore, M. Abid, and M. E. Brachet, Kolmogorov turbulence
in low-temperature superflows, Phys. Rev. Lett. \textbf{78} (1997),
3296-3299.

\bibitem{Nore97b} C. Nore, M. Abid, and M. E. Brachet, Decaying Kolmogorov
turbulence in a model of superflow, Phys. Fluids \textbf{9} (1997),
2644-2669.

\bibitem{IvCogComp} V. Ivancevic, Adaptive-Wave Alternative for the
Black-Scholes Option Pricing Model, Cogn. Comput. \textbf{2} (2010), 17--30.

\bibitem{Han} S.-H. Hanm, I.G. Koh, Stability of neural networks and
solitons of field theory. Phys. Rev. E \textbf{60}, 7608--7611, (1999)

\bibitem{LiuEtAl01} S. Liu, Z. Fu, S. Liu, Q. Zhao, Jacobi elliptic function
expansion method and periodic wave solutions of nonlinear wave equations.
Phys. Let. A \textbf{289}, 69--74, (2001)

\bibitem{LiuFan05} G-T. Liu, T-Y. Fan, New applications of developed Jacobi
elliptic function expansion methods. Phys. Let. A \textbf{345}, 161--166,
(2005)

\bibitem{AbrSte} M. Abramowitz, I.A. Stegun, (Eds): Jacobian Elliptic
Functions and Theta Functions. Chapter 16 in Handbook of Mathematical
Functions with Formulas, Graphs, and Mathematical Tables (9th ed). Dover,
New York, 567-581, (1972)

\bibitem{Yan} Z. Yan, Financial rogue waves (in press)
arXiv.q-fin.PR:0911.4259; Optical Rogue Waves (Rogons), Wolfram
Demonstration Project, (2009)

\bibitem{Solli} D.R. Solli, C. Ropers, P. Koonath, B. Jalali, Optical Rogue
Waves, Nature \textbf{450}, 1054--1057, (2007)

\bibitem{Peregrine} D.H. Peregrine, Water Waves, Nonlinear Schr\"{o}dinger
Equations and Their Solutions, J. Austral. Math. Soc. Ser. B \textbf{25},
16--43, (1983)

\bibitem{Akhmediev} N. Akhmediev, A. Ankiewicz, M. Taki, Waves That Appear
from Nowhere and Disappear without a Trace, Phys. Lett. A \textbf{373}(6),
675--678, (2009); N. Akhmediev, A. Ankiewicz, J. M. Soto-Crespo, Rogue Waves
and Rational Solutions of the Nonlinear Schr\"{o}dinger Equation, Phys. Rev.
E \textbf{80}(2), 026601, (2009)

\bibitem{Kosko1} B. Kosko, Bidirectional Associative Memory. IEEE Trans.
Sys. Man Cyb. \textbf{18}, 49--60, (1988)

\bibitem{Kosko2} B. Kosko, Neural Networks, Fuzzy Systems, A Dynamical
Systems Approach to Machine Intelligence. Prentice--Hall, New York, (1992)

\bibitem{manak74} S.V. Manakov, On the theory of two-dimensional stationary
self-focusing of electromagnetic waves. (in Russian) Zh. Eksp. Teor. Fiz.
\textbf{65}, (1973), 505-516; (transleted into English) Sov. Phys. JETP
\textbf{38}, 248--253, (1974)

\bibitem{Kerr} M. Haelterman, A.P. Sheppard, Bifurcation phenomena and
multiple soliton bound states in isotropic Kerr media. Phys. Rev. E \textbf{%
49}, 3376-3381, (1994)

\bibitem{Yang1} J. Yang, Classification of the solitary wave in coupled
nonlinear Schr\"{o}dinger equations. Physica D \textbf{108}, 92-112, (1997)

\bibitem{Benney} D.J. Benney, A.C. Newell, The propagation of nonlinear wave
envelops. J. Math. Phys. \textbf{46}, 133-139, (1967)

\bibitem{Zakharov} V.E. Zakharov, S.V. Manakov, S.P. Novikov, L.P.
Pitaevskii, Soliton theory: inverse scattering method. Nauka, Moscow, (1980)

\bibitem{Hasegawa} A. Hasegawa, Y. Kodama, Solitons in Optical
Communications. Clarendon, Oxford, (1995)

\bibitem{Radhakrishnan} R. Radhakrishnan, M. Lakshmanan, J. Hietarinta,
Inelastic collision and switching of coupled bright solitons in optical
fibers. Phys. Rev. E \textbf{56}, 2213, (1997)

\bibitem{Agrawal} G. Agrawal, Nonlinear fiber optics (3rd ed.). Academic
Press, San Diego, (2001).

\bibitem{Yang} J. Yang, Interactions of vector solitons. Phys. Rev. E
\textbf{64}, 026607, (2001)

\bibitem{Elgin} J. Elgin, V. Enolski, A. Its, Effective integration of the
nonlinear vector Schr\"{o}dinger equation. Physica D \textbf{225}(22),
127-152, (2007)

\bibitem{Lax} P. Lax, Integrals of nonlinear equations of evolution and
solitary waves. Comm. Pure Applied Math. \textbf{21}, 467--490, (1968)

\bibitem{Griffiths} D.J. Griffiths, Introduction to Quantum Mechanics (2nd
ed.), Pearson Educ. Int., (2005)

\bibitem{Thaller} B. Thaller,\ Visual Quantum Mechanics, Springer, New York,
(2000)

\bibitem{QuLeap} V. Ivancevic, T. Ivancevic, Quantum Leap: From Dirac and
Feynman, Across the Universe, to Human Body and Mind. World Scientific,
Singapore, (2008)
\end{thebibliography}
\end{document}